\documentclass{article}
\usepackage{graphicx} % Required for inserting images
\usepackage{amsmath, amssymb}
\usepackage{float}
\usepackage{cite}
\usepackage{authblk}
\usepackage{setspace}
\usepackage{xcolor}
\usepackage{titlesec}
\usepackage[rightcaption]{sidecap}

\sidecaptionvpos{figure}{c}
\usepackage[a4paper, margin=2cm]{geometry} % Adjust margin size
\usepackage{pgfplots}
\usepackage{tikz}
\usepackage{tikz-3dplot} % for 3D coordinate transformations
\usetikzlibrary{angles, quotes}
\usepackage{pgfplots} % For color gradient
\usepackage{pgfplotstable}
\usepackage{bm}
\usepackage{todonotes}
\usepgfplotslibrary{colormaps}
\usetikzlibrary{3d, decorations.pathmorphing, calc, arrows.meta}

\pgfplotsset{compat=1.18}
\usepgfplotslibrary{colormaps}
\usepackage[numbers,sort&compress]{natbib} 
\usepackage{hyperref} 
\usepackage{parskip}
\title{\textbf{\Large Radiative Signatures of Magnetic Reconnection: An Approach to Remote Probing of Reconnection Dynamics}}

\author[1,2]{Sergey~K.~Ermakov\thanks{Email: \texttt{sermakov@student.ethz.ch}}}
\author[2,3]{Michael~Bussmann}
\author[2]{Alexander~Debus}
\author[2]{Richard~Pausch}
\author[2,4]{Ulrich~Schramm}
\author[2]{René~Widera}
\author[2,3]{Klaus~Steiniger\thanks{Email: \texttt{k.steiniger@hzdr.de}}}

\affil[1]{\small Institute for Particle and Astrophysics, ETH Zürich, Wolfgang-Pauli-Strasse 27, 8093 Zürich, Switzerland}
\affil[2]{\small Helmholtz-Zentrum Dresden-Rossendorf, Bautzner Landstraße 400, 01328 Dresden, Germany}
\affil[3]{\small CASUS – Center for Advanced Systems Understanding, Untermarkt 20, 02826 Görlitz, Germany}
\affil[4]{\small Technische Universität Dresden, 01062 Dresden, Germany}

\date{}

\begin{document}

\maketitle
\noindent\rule{\linewidth}{0.4pt}

\begin{abstract}
    Magnetic reconnection drives a wide range of astrophysical phenomena, including geomagnetic storms, solar flares, and activity in blazars. However, direct measurement of key reconnection observables remains challenging due to the remote and extreme nature of these environments. While high-energy particle showers observed on Earth are often attributed to reconnection, the underlying mechanisms are not fully understood, and clear diagnostic signatures are lacking.\\
    We present a theoretical, data-driven approach for identifying reconnection radiation signatures and enabling remote diagnostics of reconnection in astrophysical settings through radiation spectra. Using particle-in-cell (PIC) simulations of magnetic reconnection, we generate radiation spectra and establish connections between spectral features and the underlying reconnection dynamics. We develop a method to estimate the ratio of the reconnection electric field to the plasmoid magnetic field from spectral data. Analytic calculations show that other parameters can be extracted in the ultra-relativistic reconnection regime, such as the magnetic field or the current sheet width.
    
\end{abstract}
\noindent\rule{\linewidth}{0.4pt}

\section{Introduction}
\label{sec:Intro}
Magnetic reconnection is a critical mechanism in astrophysics which is significant across a wide range of energetic phenomena, including geomagnetic storms~\cite{hajra2024geomagnetic}, solar winds~\cite{vemareddy2023solar}, and high-energy emissions from blazars~\cite{petropoulou2018reconnection}. While in situ missions and laboratory experiments have offered valuable insights into magnetic reconnection dynamics~\cite{hajra2024geomagnetic, ji2024low_beta_reconnection, zweibel_yamada_2010}, capturing its full scope remains challenging, since magnetic reconnection events occur in remote, inaccessible and extreme environments~\cite{petropoulou2018reconnection}.\\
Analysis of radiation from remote and extreme phenomena has become a cornerstone of astrophysical research~\cite{CORNERSTONE1, CORNERSTONE2, CORNERSTONE3, CORNERSTONE4, KHI}. In the study of magnetic reconnection, radiation measurements can provide a valuable indirect approach, especially for extreme or large environments not reproducible in laboratory setups. Recent studies have aimed to connect radiation emissions observed from blazars directly with magnetic reconnection events, highlighting a growing interest in identifying radiative signatures specific to reconnection~\cite{petropoulou2016relativistic}. Additionally, retrieving plasma parameters from radiation alone would allow for quantitative study of inaccessible reconnection events. Moreover, distinctive radiative signatures are important, as they allow to clarify the role of magnetic reconnection in a variety of astrophysical phenomena~\cite{uzdensky2011extreme}.\\
Particle-in-cell (PIC) simulations have become a widely applied method for the examination of magnetic reconnection~\cite{guo2014formation, guo2016energy, uzdensky2011extreme, pritchett2001gem}, they allow to quantitatively study the extend of magnetic reconnection in several phenomena like geomagnetic storms or solar flares as the driving mechanism. Combining PIC simulations with the analysis of radiation emission is performed in several other studies~\cite{zhang2018large, zhang2020radiationI, zhang2022radiationII, petropoulou2018reconnection}. In these works, the feasibility of relativistic magnetic reconnection as a driver for electromagnetic radiation signals of blazars is investigated, in a variety of frequency regions. Diverse radiation features are identified, most noticeably a swing in linear polarization during increasing reconnection rate and merging of plasmoids (magnetic islands filled with plasma)~\cite{zhang2018large, zhang2020radiationI, zhang2022radiationII}. The polarization swings are observed across a variety of radiation frequencies, with variation in timing. However, these works rely on a synchrotron approximation, missing out on radiation emitted by particles which are primarily accelerated by an electric field. The model averages over chunks in space, possibly missing small-scale features. A more quantitative analysis is needed, in order to extract critical observables. Finally, they focus on the highly relativistic magnetic reconnection case, which does not allow to extend their radiation observations to less energetic events such as geomagnetic storms or solar flares.\\
This study characterizes the fundamental radiative signatures of magnetic reconnection within the Harris current-sheet model. We employ a simple reconnection setup, similar to the one described by Pritchett~\cite{pritchett2001gem}, to focus on radiation from key dynamics like the increase in the reconnection electric field and the formation of power laws. It is the first fundamental study of its kind, examining radiation in detail and with regards to specific features of magnetic reconnection as never done before. Our approach identifies new radiative features and directly links them to the underlying reconnection dynamics, providing a foundational framework for future, more complex simulations. We derive a method to extract the experimentally critical reconnection-electric-field over plasmoid-magnetic-field $E/B$ from radiation data alone. Finally, we provide an outlook on quantities which can be extracted in the highly relativistic reconnection limit.\\
The paper begins with an overview of the numerical methods and a description of the magnetic reconnection dynamics observed in section~\ref{sec:Magnetic-Reconnection-Setup}. A qualitative description of the reconnection signatures is given in section~\ref{sec:Results}, enabling the derivation of experimental quantities from radiation data alone in section~\ref{sec:AnalyticModels}. The same section provides an analytic consideration on the extraction of parameters in the highly relativistic reconnection case.

%providing a plausible explanation for the power-law energy distributions frequently observed in cosmic rays~\cite{guo2014formation, guo2016energy}. 
%particularly at the extreme energy scales relevant for cosmic sources like blazars as the sources of these magnetic reconnection processes are not accessible.\\

%\section{Particle in Cell Simulations of Magnetic Reconnection}
\section{PIConGPU Simulations of Magnetic Reconnection}
\label{sec:Magnetic-Reconnection-Setup}
The simulations are performed using PIConGPU, an open-source, C++, performance-portable, and fully relativistic 3D3V particle-in-cell code~\cite{PIConGPU2013}.
PIConGPU provides a unique, in-situ radiation calculation plugin~\cite{pausch2019synthetic},
allowing computation of coherent and incoherent radiation from billions of particles while the simulation is running.

\subsection{In-Situ Radiation computation in PIConGPU}
The radiation is calculated directly via the spectrally resolved Liénard–Wiechert potential for each electron individually. The Liénard–Wiechert potentials are computed in the far field approximation, which matches the examination of astrophysical phenomena far away from the observer. The radiation is computed only for a subset of electrons in order to accelerate the computation. The full expression for such calculations is given by the following formula.
\begin{equation}\label{eq:radiation}
    \frac{d^2I}{d\Omega d\omega}=\frac{1}{16\pi^3\varepsilon_0 c}\sum_{k=1}^{N_p}q_k^2\left|\int_{t_i}^{t_f}\frac{\vec{n}\times[(\vec{n}-\vec{\beta}_k)\times\dot{\vec{\beta}}_k]}{(1-\vec{\beta}_k\cdot\vec{n})^2}e^{i\omega(t-\vec{n}\cdot\vec{r}_k/c)}\,dt\right|^2
\end{equation}
In this case, $\frac{d^2I}{d\Omega d\omega}$ denotes the spectral energy resolved by frequency $\omega$ and solid angle $\Omega$. We sum over the contributions of each of the $N_p$ macro electrons considered in the calculation. In PIC methods, electrons are grouped into macro electrons, where the position and momentum of the macro particle is evolved in time. These macro electrons represent multiple real electrons and can have different charge $q_k\neq e^-$. In the following, we refer to these macro electrons as electrons. Since the radiation is directly calculated from the macro particle dynamics, a form factor for their intrinsic charge distribution is required~\cite{pausch2019synthetic}. The subscript $k$ of a quantity signifies that it is a quantity of the $k^{\text{th}}$ electron. Each of these electrons is considered to have a position $\vec{r}_k$, velocity $\vec{\beta}_k$, and acceleration $\dot{\vec{\beta}}_k$ with a charge $q_k$. The radiation intensity measured depends on the direction in which it is measured, which is described by the normalized vector $\vec{n}$. The integral is performed from a starting time $t_i$ to a final time $t_f$. Varying $t_f$ examines the radiated energy at different times.\\
Such calculations can be accompanied by numerical artifacts. The radiation spectrum has a physical maximum frequency, beyond which no meaningful dynamics occur, called Nyquist frequency. It stems from the limited resolution of the proper timescale associated with the Nyquist frequency and therefore varies with the maximum $\gamma$ Lorentz factor. This appears as a smooth decline in radiated energy. The simulation also imposes a numerical Nyquist frequency from its discrete time steps, causing a sharp cutoff in the spectrum as in figure~\ref{fig:RadOverTime}. Second, a finite integration over the simulation box will cause a sinc-function in the radiation spectrum due to the sharp cutoff on the edges~\cite{pausch2019synthetic}. This can be avoided by applying a window function. This function decays at the edges of the box, which avoids numerical contribution from the sharp cutoff of these edges. At the same time it dampens the radiation contribution from dynamics taking place in these boundary regions. The same effect is imposed by the limited time integral from $t_i$ to $t_f$. Since the integral is taken over time, the corresponding sinc-function from the sharp cutoff appears in the frequency spectrum.

\subsection{Simulation Setup}
\begin{figure}[H]
    \centering
    \begin{minipage}{0.4\textwidth}
        \includegraphics[width=0.99\textwidth]{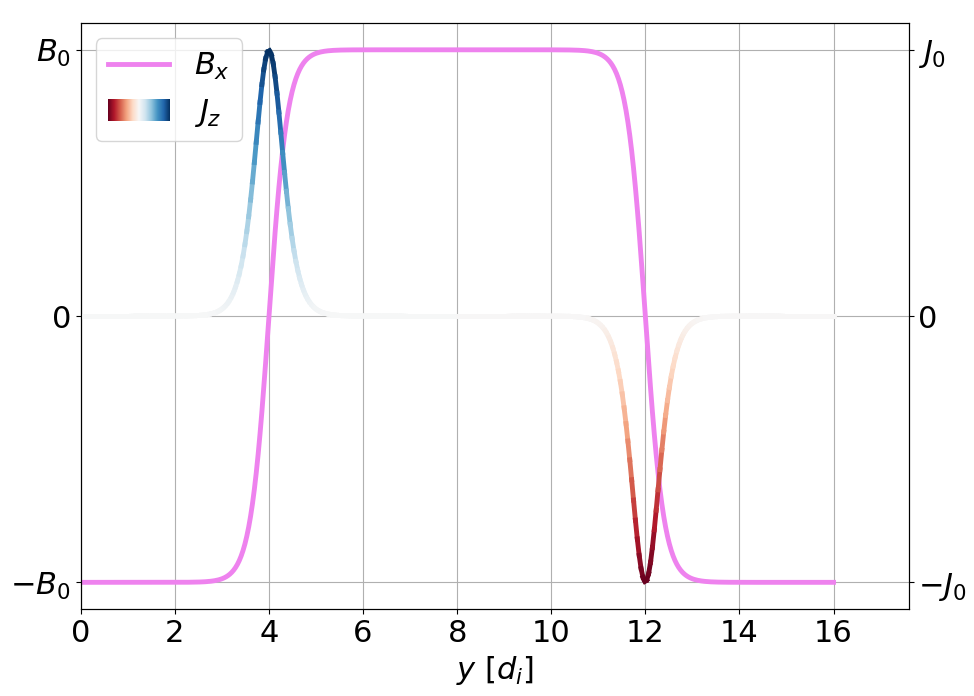}
    \end{minipage}
    \begin{minipage}{0.55\textwidth}
        \begin{tikzpicture}[line join=round, line cap=round, rotate around x=110, rotate around y=-10, rotate around z=5] % Rotate structure along x-axis
        %\begin{tikzpicture}[line join=round, line cap=round, x={(-0.8cm,-0.2cm)}, z={(0cm,1cm)}, y={(0.5cm,-0.4cm)}] % 
        
        %===================
        % Quarter Dome Grid
        %===================
        \foreach \u in {10,0,...,-100} {
            \foreach \v in {180,190,...,270} {
        
                % Fill tile (optional, just an example one)
                \ifnum \u=-40 \ifnum \v=220
                    \fill[green!70, opacity=0.8]
                    ({4*sin(\u)*sin(\v)+1}, {4*cos(\u)*sin(\v)+1.5}, {4*cos(\v)})
                    -- ({4*sin(\u+10)*sin(\v)+1}, {4*cos(\u+10)*sin(\v)+1.5}, {4*cos(\v)})
                    -- ({4*sin(\u+10)*sin(\v+10)+1}, {4*cos(\u+10)*sin(\v+10)+1.5}, {4*cos(\v+10)})
                    -- ({4*sin(\u)*sin(\v+10)+1}, {4*cos(\u)*sin(\v+10)+1.5}, {4*cos(\v+10)})
                    -- cycle;
                \fi\fi
        
                % Regular grid lines (horizontal and vertical), avoiding overflow
                \ifnum \u>-110
                    \draw[gray!90] 
                    ({4*sin(\u)*sin(\v)+1}, {4*cos(\u)*sin(\v)+1.5},{4*cos(\v)})
                    -- ({4*sin(\u+10)*sin(\v)+1}, {4*cos(\u+10)*sin(\v)+1.5},{4*cos(\v)});
                \fi
        
                \ifnum \v<270
                    \draw[gray!90] 
                    ({4*sin(\u)*sin(\v)+1}, {4*cos(\u)*sin(\v)+1.5}, {4*cos(\v)})
                    -- ({4*sin(\u)*sin(\v+10)+1}, {4*cos(\u)*sin(\v+10)+1.5}, {4*cos(\v+10)});
                \fi
        
                % Close bottom edge (last row in u)
                \ifnum \u=10
                    \ifnum \v<270
                        \draw[gray!90]
                        ({4*sin(\u)*sin(\v)+1}, {4*cos(\u)*sin(\v)+1.5}, {4*cos(\v)})
                        -- ({4*sin(\u)*sin(\v+10)+1}, {4*cos(\u)*sin(\v+10)+1.5}, {4*cos(\v+10)});
                    \fi
                \fi

            }
        }
        
        \foreach \v in {180,190,...,260} {
            \draw[gray!50]({4*sin(20)*sin(\v)+1}, {4*cos(20)*sin(\v)+1.5}, {4*cos(\v)})
            -- ({4*sin(20)*sin(\v+10)+1}, {4*cos(20)*sin(\v+10)+1.5}, {4*cos(\v+10)});
        }

        %===================
        % Cuboids Stacked Along the y-Axis (Half Length)
        %===================
        % Define Color Gradient Function (based on e^y)
        % Define a color function based on exp(y) 0, 0.05, 0.1, ..., 2.5
        %\foreach \y in {0, 0.05, 0.1, ..., 2.5} 
        \foreach \i in {0,...,50}
        {
            \pgfmathsetmacro{\y}{0.05*\i}
        
            % Red component (centered near 0.3*2.5)
            \pgfmathsetmacro{\separateVarOne}{min((exp(min((\y-0.3*2.5)/(0.2),4)) + exp(min(-(\y-0.3*2.5)/(0.2),4)))^2, 1000)}
            \pgfmathsetmacro{\redIntensityRaw}{max(0, min(1.0 - 4/\separateVarOne, 1))}
        
            % Blue component (centered near 0.7*2.5)
            \pgfmathsetmacro{\separateVarTwo}{min((exp(min((\y-0.7*2.5)/(0.2),4)) + exp(min(-(\y-0.7*2.5)/(0.2),4)))^2, 1000)}
            \pgfmathsetmacro{\blueIntensityRaw}{max(0, min(1.0 - 4/\separateVarTwo, 1))}
        
            % Suppress overlaps to avoid magenta
            \pgfmathsetmacro{\redIntensity}{\redIntensityRaw * (1 - \blueIntensityRaw)}
            \pgfmathsetmacro{\blueIntensity}{\blueIntensityRaw * (1 - \redIntensityRaw)}
        
            % White background: make green same as red/blue base level
            % Essentially: 1 when both R and B are low
            \pgfmathsetmacro{\whiteLevel}{max(0, 1 - (\redIntensity + \blueIntensity))}
            \pgfmathsetmacro{\greenIntensity}{\whiteLevel}
            \pgfmathsetmacro{\redIntensity}{\redIntensity + \whiteLevel}
            \pgfmathsetmacro{\blueIntensity}{\blueIntensity + \whiteLevel}
        
            % Clip to [0,1]
            \pgfmathsetmacro{\redIntensity}{min(1, \redIntensity)}
            \pgfmathsetmacro{\greenIntensity}{min(1, \greenIntensity)}
            \pgfmathsetmacro{\blueIntensity}{min(1, \blueIntensity)}
        
            % Draw faces
            \ifnum \i < 25
                \filldraw[fill={rgb,1:red,\redIntensity;green,\greenIntensity;blue,\blueIntensity}, draw=none]
                    (-3,\y,-0.5) -- (3,\y,-0.5) -- (3,\y,0.5) -- (-3,\y,0.5) -- cycle; % Bottom Face
            \else
                \filldraw[fill={rgb,1:red,\redIntensity;green,\greenIntensity;blue,\blueIntensity}, draw=none]
                    (-3,\y,-0.5) -- (3,\y,-0.5) -- (3,\y,0.5) -- (-3,\y,0.5) -- cycle; % Bottom Face
            \fi
        
            \filldraw[fill={rgb,1:red,\redIntensity;green,\greenIntensity;blue,\blueIntensity}, draw=none]
                (-3,\y+0.02,-0.5) -- (3,\y+0.02,-0.5) -- (3,\y+0.02,0.5) -- (-3,\y+0.05,0.5) -- cycle; % Top Face
        
            \filldraw[fill={rgb,1:red,\redIntensity;green,\greenIntensity;blue,\blueIntensity}, draw=none]
                (3,\y,-0.5) -- (3,\y+0.02,-0.5) -- (3,\y+0.02,0.5) -- (3,\y,0.5) -- cycle; % Right Face
        }

        % Draw Front and Back Faces to fully enclose the box
        %\filldraw[black!20] (-3,0,-0.5) -- (-3,2.5,-0.5) -- (3,2.5,-0.5) -- (3,0,-0.5) -- cycle; % Front Face
        %\filldraw[black!30] (-3,0,0.5) -- (-3,2.5,0.5) -- (3,2.5,0.5) -- (3,0,0.5) -- cycle; % Back Face
        
        % Axes for reference
        %\draw[thick,->] (-4,1,0.5) -- (-4,-0.5,0.5) node[above] {$y$};  % y-axis
        %\draw[thick,->] (-4,1,0.5) -- (-2.5,1,0.5) node[right] {$x$};  % x-axis
        %\draw[thick,->] (-4,1,0.5) -- (-4,1,-1) node[below left] {$z$}; % z-axis

        \coordinate (A) at (-3,0,-0.5); 
        \coordinate (B) at (3,0,-0.5); 
        \coordinate (C) at (3,0,0.5);
        \coordinate (D) at (-3,0,0.5);
        \coordinate (E) at (-3,0,-0.5);
        \coordinate (F) at (3,0,-0.5);
        \coordinate (G) at (3,0,0.5);
        \coordinate (H) at (-3,0,0.5);
        \coordinate (I) at (-3,0,-0.5); 
        \coordinate (J) at (-3,2.5,-0.5);
        \coordinate (K) at (3,2.5,-0.5);
        \coordinate (L) at (3,0,-0.5);
        \coordinate (M) at (-3,0,0.5);
        \coordinate (N) at (-3,2.5,0.5);
        \coordinate (O) at (3,2.5,0.5);
        \coordinate (P) at (3,0,0.5);
        
        \draw[thin] (I) -- (J) -- (K) -- (L) -- cycle;
        \draw[thin] (J) -- (N) -- (O) -- (K);
        \draw[thin] (O) -- (C) -- (L);

        % Upward Arrow on First Red Cuboid
        \draw[thick, ->, black] (3, 0.75, 0.4) -- (3, 0.75, -0.4);
        % Upward Arrow on First Red Cuboid
        \draw[thick, ->, black] (3, 1.75, -0.4) -- (3, 1.75, 0.4) node[below]{$\vec{J}$};
        
        %\foreach \y in {0, 0.25, 0.5, 0.75, 1.0, 1.25, 1.5, 1.75, 2.0, 2.25} {
        %    \draw[thin, black, opacity=0.6] (-3,\y,-0.4) -- (2.8,\y,-0.4);
        %}
        \pgfmathsetmacro{\y}{0};
        \draw[thick, purple!60!black, opacity=0.7] (2.0,\y,-0.4) -- (2.8,\y,-0.4);
        \draw[thick, ->, >=stealth, purple!60!black, opacity=0.7] (-3,\y,-0.4) -- (2.0,\y,-0.4);
        \pgfmathsetmacro{\y}{0.25};
        \draw[thick, purple!60!black, opacity=0.7] (2.0,\y,-0.4) -- (2.8,\y,-0.4);
        \draw[thick, ->, >=stealth, purple!60!black, opacity=0.7] (-3,\y,-0.4) -- (2.0,\y,-0.4);
        \pgfmathsetmacro{\y}{0.5};
        \draw[thick, purple!60!black, opacity=0.7] (2.0,\y,-0.4) -- (2.8,\y,-0.4);
        \draw[thick, ->, >=stealth, purple!60!black, opacity=0.7] (-3,\y,-0.4) -- (2.0,\y,-0.4);
        \pgfmathsetmacro{\y}{0.75};
        \draw[thick, purple!60!black, opacity=0.7] (2.0,\y,-0.4) -- (2.8,\y,-0.4);
        \draw[thick, ->, >=stealth, purple!60!black, opacity=0.7] (-3,\y,-0.4) -- (2.0,\y,-0.4);
        \pgfmathsetmacro{\y}{1.0};
        \draw[thick, purple!60!black, opacity=0.7] (-3,\y,-0.4) -- (1.8,\y,-0.4);
        \draw[thick, ->, >=stealth, purple!60!black, opacity=0.7] (2.8,\y,-0.4) -- (1.8,\y,-0.4);
        \pgfmathsetmacro{\y}{1.25};
        \draw[thick, purple!60!black, opacity=0.7] (-3,\y,-0.4) -- (1.8,\y,-0.4);
        \draw[thick, ->, >=stealth, purple!60!black, opacity=0.7] (2.8,\y,-0.4) -- (1.8,\y,-0.4);
        \pgfmathsetmacro{\y}{1.5};
        \draw[thick, purple!60!black, opacity=0.7] (-3,\y,-0.4) -- (1.8,\y,-0.4);
        \draw[thick, ->, >=stealth, purple!60!black, opacity=0.7] (2.8,\y,-0.4) -- (1.8,\y,-0.4);
        \pgfmathsetmacro{\y}{1.75};
        \draw[thick, purple!60!black, opacity=0.7] (2.0,\y,-0.4) -- (2.8,\y,-0.4);
        \draw[thick, ->, >=stealth, purple!60!black, opacity=0.7] (-3,\y,-0.4) -- (2.0,\y,-0.4);
        \pgfmathsetmacro{\y}{2.0};
        \draw[thick, purple!60!black, opacity=0.7] (2.0,\y,-0.4) -- (2.8,\y,-0.4);
        \draw[thick, ->, >=stealth, purple!60!black, opacity=0.7] (-3,\y,-0.4) -- (2.0,\y,-0.4);
        \pgfmathsetmacro{\y}{2.25};
        \draw[thick, purple!60!black, opacity=0.7] (2.0,\y,-0.4) -- (2.8,\y,-0.4) node[midway, below, yshift=-2pt] {$\vec{B}$};
        \draw[thick, ->, >=stealth, purple!60!black, opacity=0.7] (-3,\y,-0.4) -- (2.0,\y,-0.4);
        %===================
        % Wavy Waves from First Red Cuboid
        %===================
        % Red Wave
        \draw[decorate, decoration={snake, amplitude=1mm, segment length=3mm}, thick, green]
            (0,0.75,-0.4) -- ({4*sin(-35)*sin(225)+1}, {4*cos(-35)*sin(225)+1.5},{4*cos(225)});
        \draw[thin, black, dashed]
            (0,0.75,-0.4) -- ({4*sin(-40)*sin(220)+1}, {4*cos(-40)*sin(220)+1.5}, {-0.4});
        \draw[thin, black, dashed]
            ({4*sin(-40)*sin(220)+1}, {4*cos(-40)*sin(220)+1.5}, {-0.4}) -- ({4*sin(-40)*sin(220)+1}, {4*cos(-40)*sin(220)+1.5}, {4*cos(220)});
        
        % green Wave
        %\draw[decorate, decoration={snake, amplitude=1mm, segment length=3mm}, thick, green]
        %    (0,0.75,-0.4) -- ({4*sin(-15)*sin(205)+1}, {4*cos(-15)*sin(205)+1.5},{4*cos(205)});

        \draw[thin, black, dashed, ->]
            (0,0.75,-0.4) -- (0,0.75-1.5,-0.4)node[above] {$y$};
        \draw[thin, black, dashed, ->]
            (0,0.75,-0.4) -- (1.5,0.75,-0.4)node[below] {$x$};
        \draw[thin, black, dashed, ->]
            (0,0.75,-0.4) -- (0,0.75,-1.5-0.4)node[above] {$z$};
        
        \draw[thick, black, ->]
            (-3.4,1.5,0.5) -- (-0.1,0.75+0.1,-0.4) node[midway, above, black] {$\vec{r}$};

        \coordinate (A) at (1.5,0.75,-0.4);
        \coordinate (B) at (0,0.75,-0.4);
        \coordinate (C) at ({4*sin(-40)*sin(220)+1}, {4*cos(-40)*sin(220)+1.5}, {-0.4});
        \coordinate (D) at ({4*sin(-40)*sin(220)+1}, {4*cos(-40)*sin(220)+1.5},{4*cos(220)});
        \coordinate (E) at (0,0.75,-1.5-0.4);
        % Draw the angle marker (with label in black)
        \pic [draw, "$\bm{\phi}$", angle eccentricity=0.90, angle radius=1.2cm] 
          {angle = A--B--C};
        
        \pic [draw, "$\bm{\theta}$", angle eccentricity=0.75, angle radius=0.82cm] 
          {angle = D--B--E};

        % Green Wave
        %\draw[decorate, decoration={snake, amplitude=1mm, segment length=3mm}, thick, green]
        %    (0,0.75,-0.4) -- ({4*sin(-95)*sin(255)+1}, {4*cos(-95)*sin(255)+1.5},{4*cos(255)});
        \filldraw[black] (-0.05,0.75,-0.4) circle (2.5pt) node[below, yshift=1.5pt, black] {$e^-$};

        %===================
        % Coordinate Axes
        %===================
        \draw[thick,->] (-3.4,1.5,0.5) -- (-3.4,0.5,0.5) node[above] {$y$};  % y-axis
        \draw[thick,->] (-3.4,1.5,0.5) -- (-2.4,1.5,0.5) node[below] {$x$};  % x-axis
        \draw[thick,->] (-3.4,1.5,0.5) -- (-3.4,1.5,-0.5) node[left] {$z$}; % z-axis
        
        %\draw[thin, grey, domain=0:2.5, samples=100] plot (\x, {tanh(\x)}) node[right] {$z = \tanh(y)$};
        
        \end{tikzpicture}
        
    \end{minipage}
    \caption{(Left) Initial magnetic field \( B_x \) and current sheet density \( J_z \) (equation~\ref{eq:Harris}). The plasma current density curve is color-coded by magnitude, matching the scale used in the right panel. (Right) Schematic of the simulation setup. Purple lines show initial magnetic field lines; the colormap shows initial charge current density. An electron (charge \( e^- \)) at position \( \vec{r} \) emits radiation detected at polar angle \( \phi \) and azimuthal angle \( \theta \) by a pixelated segment of the surrounding radiation half-dome.}
    \label{fig:Schematic}
\end{figure}
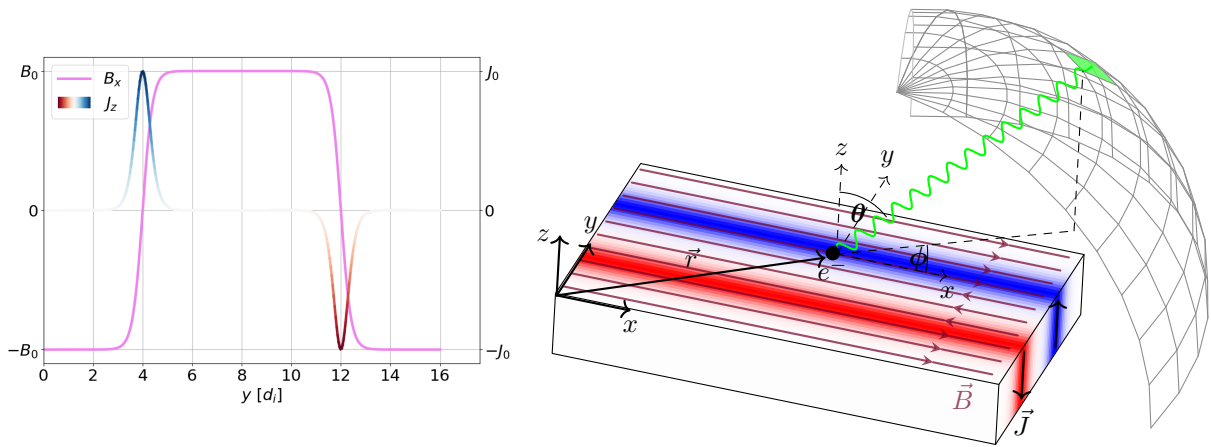

A basic simulation setup is chosen, to focus on the essential dynamics of magnetic reconnection. The setup is inspired by the Pritchett GEM-challenge paper~\cite{pritchett2001gem}, which is known to have well-understood and characteristic dynamics for magnetic reconnection. Initially, a Harris current sheet is applied, which is a charge flux modeled closely to experimental observations of magnetic reconnection~\cite{harris1962plasma}. This current sheet is a steady-state solution of the MHD equations with a corresponding magnetic field, where both current sheet and magnetic field are given by 
\begin{equation}\label{eq:Harris}
    %\centering
    \begin{split}
        \vec{J}(\vec{r})=J_0\,\text{sech}^2\left(\frac{y}{\lambda}\right)\vec{e_z}\\
        \vec{B}(\vec{r})=B_0\,\text{tanh}\left(\frac{y}{\lambda}\right)\vec{e_x}
    \end{split}    
\end{equation}
Here, $J_0=n_0e(\gamma_i\beta_i+\gamma_e\beta_e)c$ is the initial charge current density, where $n_0$, $\gamma_i\beta_i+\gamma_e\beta_e$, and $B_0$ denote the maximum plasma density, initial velocity of ions plus electrons, boosted by the according Lorentz factors, and amplitude of the magnetic field, respectively. For magnetic reconnection dynamics to be initialized, a small perturbation is necessary, since the solution provided by equation~\ref{eq:Harris} is stationary. Often, it is given by a perturbative wave of the form in equation~\ref{eq:Perturbation}~\cite{li_effect_2021, pritchett2001gem}.
\begin{equation}\label{eq:Perturbation}
    \delta\vec{B}(\vec{r})=\delta B_0k_y\,\cos(k_xx)\sin(k_yy)\vec{e_x}-\delta B_0k_x\,\sin(k_xx)\cos(k_yy)\vec{e_y}
\end{equation}
The magnetic field strength of the perturbation is denoted by $\delta B_0$ and $k_x=\frac{2\pi}{L_x}$, $k_y=\frac{\pi}{L_y}$ with $L_x$ and $L_y$ the numerical box size in $x$- and $y$-directions respectively. The perturbation term is shifted in this setup by a quarter of the simulation domain, such that both the $X$-shaped and the $O$-shaped magnetic field lines lie at the edges of the box ($X$-shaped and $O$-shaped magnetic field lines are highlighted in figure~\ref{fig:EZ}). These two structures are equally relevant for characteristic dynamics of magnetic reconnection and should therefore contribute to the radiation with equal strength. As the window function of the radiation plugin decays at the boundaries of the simulation box, we ensure both structures experience the decay of the window function equally.\\
Parameters can resemble the near earth and sun environment~\cite{cassak_review_2017, 1mTB}, but some quantities are scaled down in order to enable reasonable simulation time scales. The initial magnetic field is $B_0 = 1$~mT with a perturbation term of 1\%. An ion temperature of 19.1~keV is applied. A temperature ratio of $T_e/T_i=0.2$ and a mass ratio of $m_i/m_e=25$ is chosen. The current sheet thickness is $\lambda=1.25d_i$, where $d_i=\frac{c}{\omega_{pi}}$, with $\omega_{pi}$ the ion plasma frequency, is the ion inertial skin depth. A double current sheet configuration is chosen with periodic boundary conditions in order to avoid any numerical impact of non-periodic boundary conditions (such as reflective) on both radiation and physics.\\
The radiation depends on the observation angle and is therefore calculated from multiple observation angles. A quarter dome, as shown in figure~\ref{fig:Schematic}, which pixelated observation directions, is employed to capture the directionality of radiation. A quarter is sufficient to capture all radiative features due to the symmetry of the setup. Additionally, only the radiation of the electrons from the lower current sheet is calculated in order to avoid interference of radiation of the two different reconnection events.
The setup employs a $2048 \times 1024$ cell grid with each cell of size 2~cm$\times$2~cm. The electromagnetic fields are simulated via a Yee solver~\cite{yee1966numerical}, and particles are propagated with a Higuera–Cary Pusher~\cite{higuera2017structure}. They influence the fields through an Esirkepov current deposition scheme, where macroparticles take a piecewise cubic-spline shape~\cite{esirkepov2001exact}.

\subsection{Magnetic Reconnection Dynamics}
A multitude of processes characteristic for magnetic reconnection is observed. For the consideration of radiation signals, we divide relevant dynamics into three groups: $1^{\text{st}}$ the initial dynamics, $2^{\text{nd}}$ the electric field dominated dynamics and $3^{\text{rd}}$ the magnetic field dominated dynamics.

\subsubsection{Initial Dynamics}
\begin{figure}[H]
    \centering
    \includegraphics[width=0.99\textwidth]{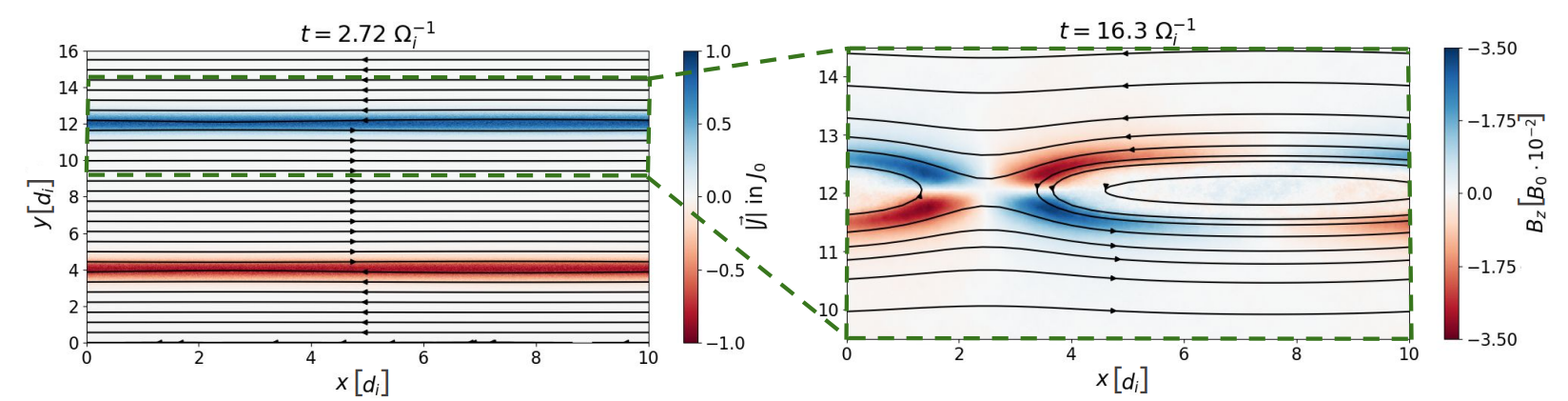}
    \caption{(Left) Charge current density in this setup at time $t=2.72\,\Omega_i^{-1}$. The initially dominant out-of-plane component is shown. (Right) Out-of-plane Hall magnetic field at time $t=16.3\,\Omega_i^{-1}$.}
    \label{fig:JPlots}
\end{figure}

The initial dynamics is characterized by the double current sheet configuration shown in figure~\ref{fig:JPlots}. This corresponds to the steady state solutions in equation~\ref{eq:Harris} as discussed before. In subsequent figures, only the lower current sheet is displayed, as radiation is sampled exclusively from this sheet. All times are given in multiples of the inverse ion cyclotron frequency $\Omega_i^{-1}$ and all lengths are given in multiples of the ion inertial skin depth $d_i$.
After the perturbation as in equation~\ref{eq:Perturbation}, the setup shows Hall magnetic fields, which we include in our initial dynamics classification. The Hall magnetic field is a quadrupole out-of-plane magnetic field structure, originating from the circular movement of electrons around the reconnection region. It is clearly visible in the red and blue quadrupole like structure of the color map of the out-of-plane magnetic field in the right part of figure~\ref{fig:JPlots}.

\subsubsection{Electric Field Dominated Dynamics}
\begin{figure}[H]
    \centering
    \includegraphics[width=0.99\textwidth]{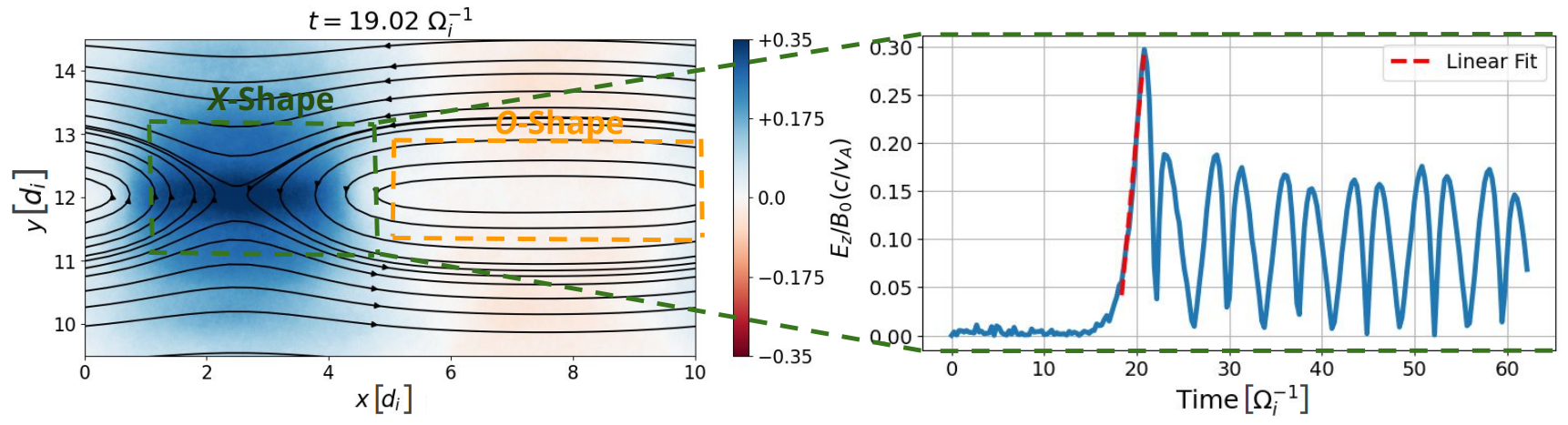}
    \caption{(Left) Out-of-plane electric field due to rising reconnection rate at time $t=19.02\,\Omega_i^{-1}$. The green box highlights the region of the $X$-shaped magnetic field lines. The change in time of these magnetic field lines are the cause of the rising out-of-plane electric field. On the right side of the green box, the magnetic field lines form an $O$-shape. (Right) Out-of-plane electric field averaged across the area of the green box on the left. This indicates the time relevant for electric field dynamics is between 18~$\Omega_i^{-1}$ and 20~$\Omega_i^{-1}$. In this time frame, the increase in electric field over time is approximately linear (indicated by the red dashed line).}
    \label{fig:EZ}
\end{figure}

\begin{figure}[H]
    \centering
    \includegraphics[width=0.99\textwidth]{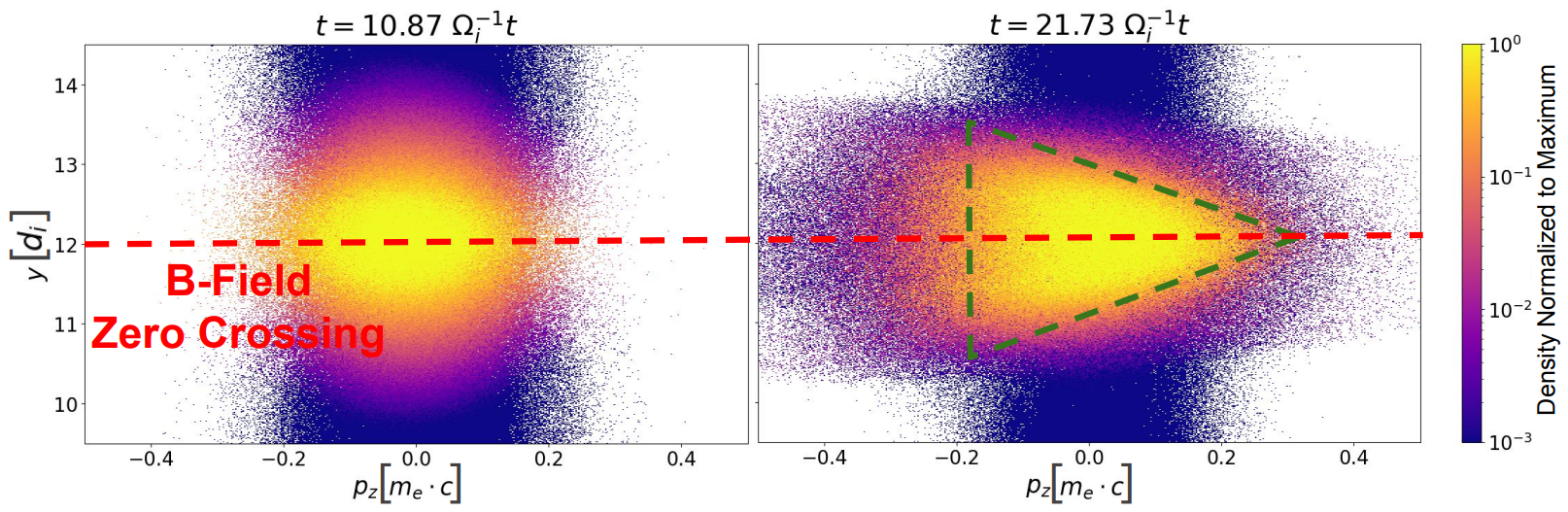}
    \caption{
    %(Left) Plasmoids after the initial reconnection event at $t=54.99\,\Omega_i^{-1}$. 
    2D Phasespace plot of $y$ coordinate and out-of-plane momentum $p_z$. The red dashed line indicates the position of the zero crossing, where the initial magnetic field is equal to zero. The initially symmetric density evolves into a triangular shape, exceeding the initial maximum momentum of $0.6m_ec$.}
    \label{fig:PSDensity}
\end{figure}

We consider the second stage next, which we labeled as the electric field dominated dynamics. It is characterized by the rise in the out-of-plane electric field, which can be seen in figure~\ref{fig:EZ}. The out of plane electric field is induced by the rapidly changing magnetic field in the $X$-magnetic field line shaped region (highlighted by the green box in figure~\ref{fig:EZ}), called the reconnection region~\cite{zenitani2001generation}. This is shown by the strong blue color in the color map of the out-of-plane electric field in figure~\ref{fig:EZ}.
The averaged electric field in the region of reconnection is displayed over time. This allows to quantify the time period, in which this electric field dominated dynamics are most relevant. Figure~\ref{fig:EZ} on the right shows clearly that a sharp increase in electric field happens between 18$~\Omega_i^{-1}$ and 20$~\Omega_i^{-1}$. It is evident that a linear increase in electric field is a good approximation for this period.\\
We examine the increase of particle energy during this process. Consider the 2D phase space in figure~\ref{fig:PSDensity} of $y$-coordinate (on the $y$-axis) and the out-of-plane $p_z$ component (on the $x$-axis). There is a high density of electrons near the zero crossing of the initial magnetic field configuration (shown on the left of figure~\ref{fig:Schematic}). Additionally, the acquisition in high momentum during the time frame of dominant electric field dynamics (figure~\ref{fig:EZ} shows the electric field evolution over time). The phase space density acquires a triangular shape, indicated by the green dashed lines in figure~\ref{fig:PSDensity}. This reflects the dynamics induced by the out-of-plane electric field: electrons near the magnetic field's zero crossing gain high positive out-of-plane momentum due to minimal magnetic influence. The triangular shape directly illustrates this, highlighting that momentum is acquired by particles near the field zero-crossing.
%This quantity is also used as a verification of our simulation, by performing a simulation with identical parameters as those by Pritchett~\cite{pritchett2001gem} (right plot, figure~\ref{fig:EoverTime}), which shows good agreement. Notice that in our setup, the rising reconnection rate resembles the same shape as in Pritchett's simulation, which is well approximated by a linear increase in electric field (left plot, figure~\ref{fig:EoverTime}).
%\begin{figure}[H]
%    \centering
    %\includegraphics[width=0.49\textwidth]{plots/ErateOverTime.png}
%    \caption{Out-of-plane electric field over time. (Left) Simulation of setup with parameters identical to Pritchett. (Right) Out-of-plane electric field from the setup employed in this paper. This indicates the time relevant for electric field dynamics is between 18$\Omega_i^{-1}$ and 20$\Omega_i^{-1}$.}
%    \label{fig:EoverTime}
%\end{figure}

\subsubsection{Magnetic Field Dominated Dynamics}\label{sec:MagneticFieldDynamics}
\begin{SCfigure}[50][b!]
    \centering
    \includegraphics[width=0.5\textwidth]{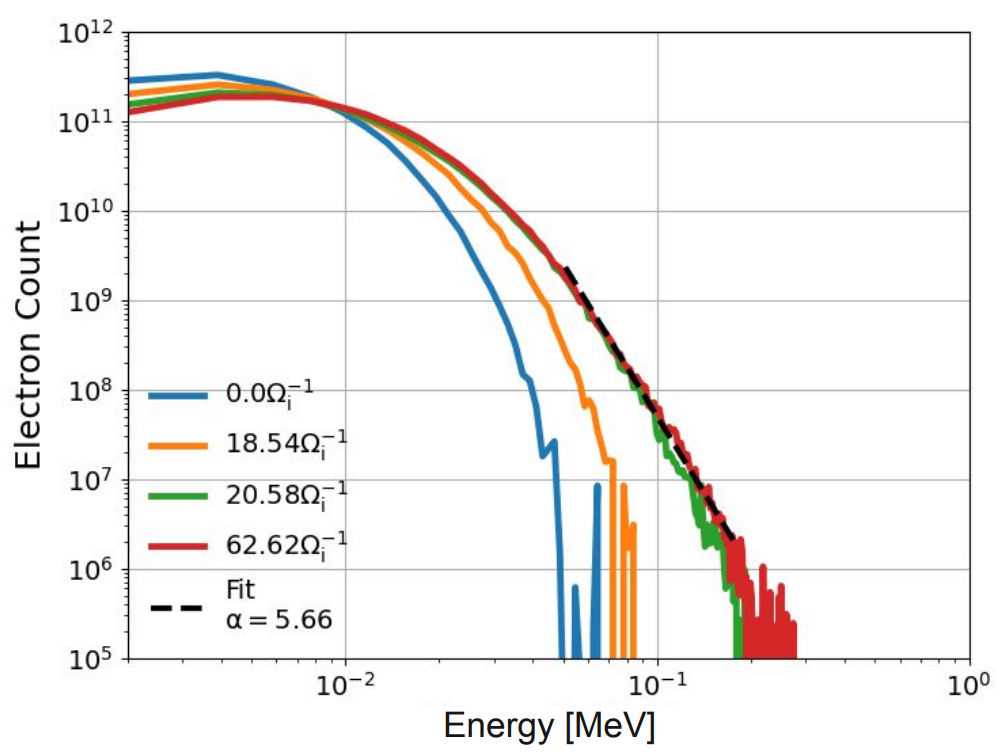}
    \caption{Energy histograms showing formation of power-law energy spectra over time. In this case, the power-law is governed by $E^{-\alpha}$ with $\alpha=5.66$. Notice the energy increase evolving into a power law during the electric field dominant period of $18~\Omega_i^{-1}-20~\Omega_i^{-1}$. The energy increase proceeds beyond this time frame, as indicated in the subsequent figure~\ref{fig:PlasmoidOscillationEnergy}.}
    \label{fig:PowerLaw}
\end{SCfigure}

\begin{figure}[H]
    \centering
    \includegraphics[width=0.99\textwidth]{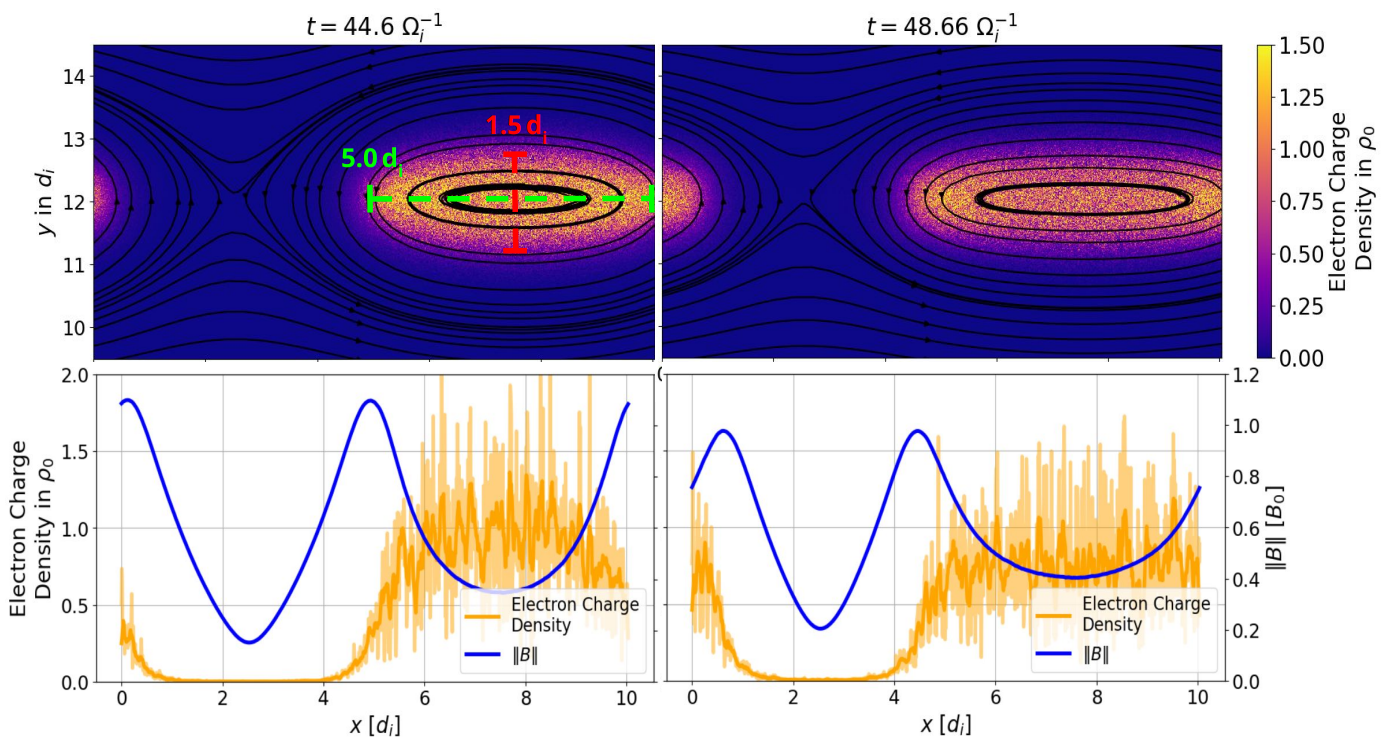}
    \caption{(Top) Plasmoids at different points in time show the oscillatory behavior of the semi-major and semi-minor axes of their elliptical magnetic field structure. While the semi-major axis shows a strong change (marked in green), the semi-minor axis shows less variation (marked in red). (Bottom) An $x$-outline of the variation in electric charge density of the electrons and absolute magnetic field $\|B\|$. The orange curve is a denoised charge density curve, while the light orange background is the charge density before denoising. $\rho_0$ denotes the charge density at the beginning of the simulation. Notice that most electrons towards the middle of the plasmoid experience a small variation in magnetic field. When coming closer to the edges of the plasmoid, where the electric charge density drops, the magnetic field sharply increases. Since a high density of electrons gathers in the middle of the plasmoid we can approximate that the electrons experience a magnetic field on the order of $\sim0.5B_0$.}
    \label{fig:PlasmoidOscillation}
\end{figure}

\begin{figure}[H]
    \centering
    \includegraphics[width=0.44\textwidth]{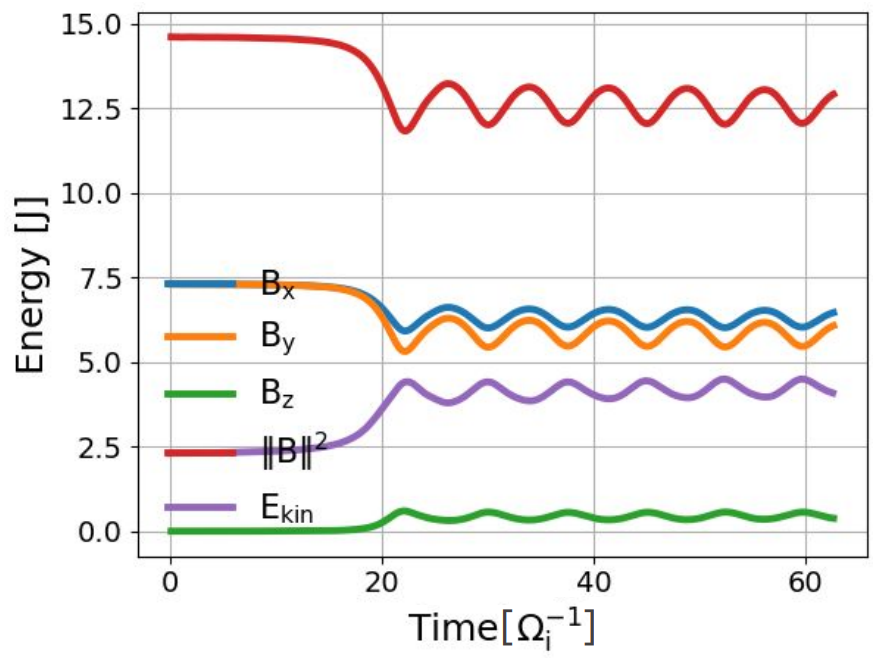}
    \includegraphics[width=0.54\textwidth]{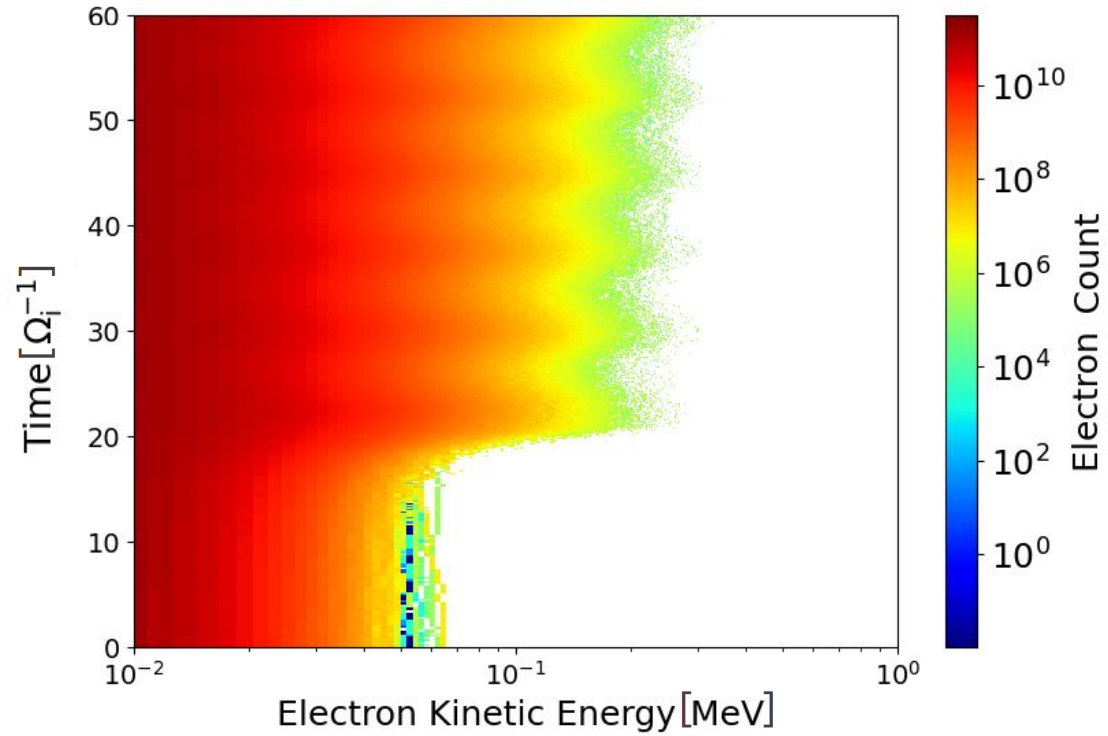}
    \caption{(Left) Magnetic energy plotted together with the kinetic energy of all particles. An energy conversion process is visible, where kinetic energy is periodically converted into magnetic energy. (Right) A 2D histogram of the energies of electrons used for radiation calculations. It shows the same oscillatory behavior as the magnetic field energy. This plot also reveals the strong correlation between electron energy and oscillating spectral energy in figure~\ref{fig:RadOverTime}.}
    \label{fig:PlasmoidOscillationEnergy}
\end{figure}

The third stage, the magnetic field dominated dynamics, is characterized by the electrons trapped in plasmoids. Plasmoids are magnetic islands with the electrons previously accelerated by the electric field trapped inside~\cite{guo2014formation} (figure~\ref{fig:PlasmoidOscillation} shows the evolution of these plasmoids over time) The processes involving plasmoids are characterized by the time period beyond 20~$\Omega_i^{-1}$.\\ 
The electrons trapped inside the plasmoids follow a power-law energy spectrum~\cite{guo2014formation, guo2016energy}, where the time evolution of this energy spectrum is shown in figure~\ref{fig:PowerLaw}. The greatest increase in high energy electrons happens over the same period of time (between 18~$\Omega_i^{-1}$ and 20~$\Omega_i^{-1}$) as the electric field acceleration, discussed in the second stage and depicted in figure~\ref{fig:EZ}. 
%\begin{figure}[H]
%    \centering
    %\includegraphics[width=0.549\textwidth]{plots/plasmoidFinal.png}
%    \includegraphics[width=0.55\textwidth]{plotsNew/EnergyHistogram.png}
%    \caption{
    %(Left) Plasmoids after the initial reconnection event at $t=54.99\,\Omega_i^{-1}$. 
%    Energy histograms showing formation of power-law energy spectra over time. In this case, the power-law is governed by $E^{-\alpha}$ with $\alpha=5.66$.}
%    \label{fig:PowerLaw}
%\end{figure}
The plasmoids constitute the electrons following a power-law energy distribution. This plasmoid exhibits a steady-state oscillations for the remaining duration of the simulation. Figure~\ref{fig:PlasmoidOscillation} shows these elliptical plasmoids oscillating via periodic changes in their semi-major (green dashed) and semi-minor (red dashed) axes.\\
Figure~\ref{fig:PlasmoidOscillationEnergy} reveals matching oscillations in both kinetic and magnetic energies, reflected in the integrated energies (left plot) and the 2D electron energy histogram (right plot).\\
This suggests that the plasmoids undergo a harmonic-oscillator-like process, converting the initially obtained kinetic energy into magnetic energy of the plasmoids periodically. This is expected to reflect strongly in the calculated radiated energy and radiation polarization.

\subsubsection{Transition from Electric- to Magnetic-Field Dynamics}\label{sec:EtoMag}
Figure~\ref{fig:PlasmoidOscillationEnergy} shows the electron energy evolution over time, defining the key time frames for analysis. While the electric field increase happens between $18~\Omega_i^{-1}$ and $20~\Omega_i^{-1}$, the particle energy increase persists until $22~\Omega_i^{-1}$. This is the time frame relevant for radiation analysis, since we expect the strong acceleration of electrons by the electric field to result in an increase in radiated energy.\\
In the magnetic-dominated regime, any time frame post electron acceleration by the electric field can be selected. Thus, when considering magnetic field dynamics, we examine times beyond $22~\Omega_i^{-1}$. Past this point in time, any time frame can be considered for magnetic field dynamics, since the setup remains in the periodic plasmoid-oscillation relaxation-state. The definition of these time frames allows to examine the radiation of magnetic reconnection in the context of the dynamics present at the moment of radiation. This allows to uniquely identify the cause for signals in the observed spectra. 
The individual features discussed in this section are relevant for the radiation signals observed. They all leave distinct signatures, which in some cases allow to deduce physical quantities from radiation data alone.

\section{Characteristic Radiation Features for Magnetic Reconnection}
\label{sec:Results}
\begin{figure}[H]
    \centering
    \includegraphics[width=0.99\textwidth]{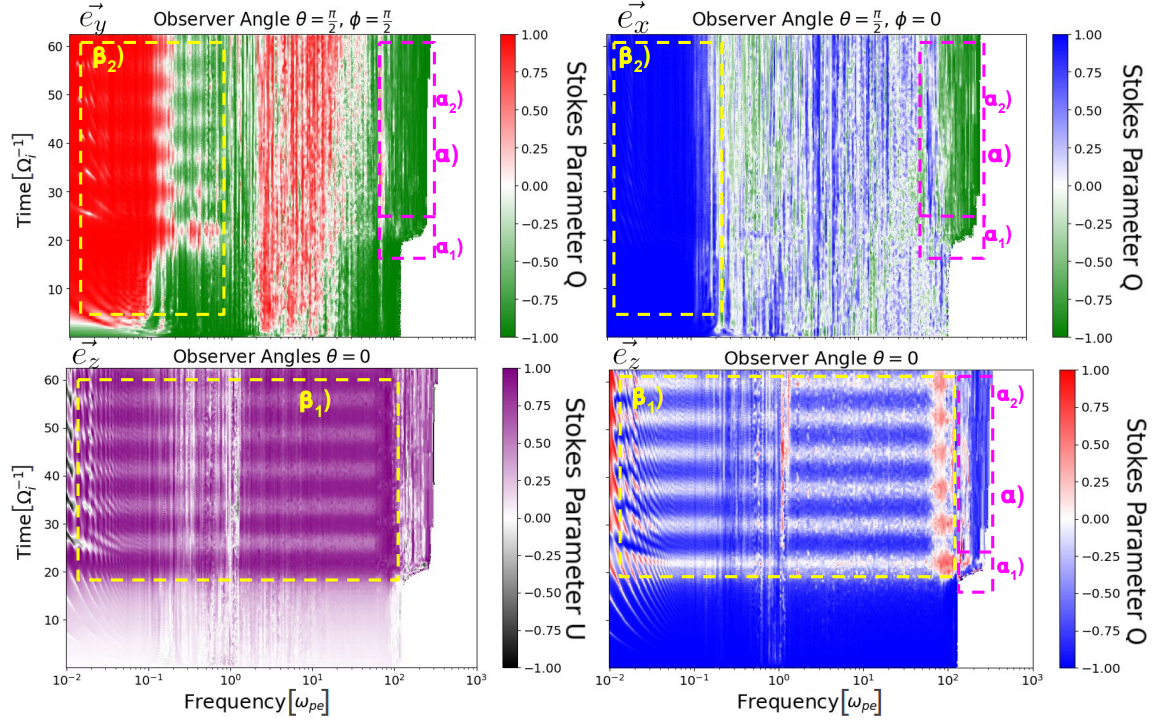}
    \caption{Polarization over time depicted by normalized Stokes parameters. The relevant signals include: $\alpha)$ polarized radiation from the electrons after initial electric field acceleration in the magnetic field. $\beta), \beta_1)$ and $\beta_2)$ denote the polarization produced by the oscillating plasmoid.\\ 
    The colors indicate the polarization of the radiation. Since we observe from all three orthogonal directions of space, the polarization directions of the electromagnetic radiation corresponds to the directions in space. The color red indicates $x$-polarized, the color blue indicates $y$-polarized and the color green indicates $z$-polarized radiation. Purple indicates the diagonally polarized radiation between the $x$- and the $y$-axis and black indicates the component orthogonal to it. Stokes parameters not shown here either show a superposition of the signals observed here or show limited to no polarization.}
    \label{fig:PolarizationOverTime}
\end{figure}

%\subsection{Pritchett Setup}
%\label{sec:Pritchett-Setup}
%\input{Pritchett-Setup.tex}
\label{sec:Time-Evolution-of-Frequency-Spectra}
%\todo{Verweise bei Erklaerungen auf vorher gezeigte Partikeldynamik}
The field and particle dynamics observed in the magnetic reconnection simulation reflect in the simulated electron radiation. This can later be exploited for the extraction of the reconnection-rate from radiation data alone. Here we provide an overview of the typically observed radiation features.\\
Typically, radiation is displayed as intensity versus frequency spectrum on a 1D graph. However, integrating the frequency spectrum over the entire time domain is not practical, as it would blend signals from different reconnection events occurring at various times. Instead, the approach here involves plotting the cumulative integral of the signal over time (indicated by equation~\ref{eq:radiation}), gradually extending the integration period. This method captures the evolution of radiation spectra at each moment, allowing for clearer identification of individual events and their characteristics throughout the reconnection process. This, however, comes at a cost. The integral in equation~\ref{eq:radiation} computes over a limited period in time, resulting in a similar effect as the finite space of the simulation box. The finite integral results in a sinc-function as discussed in section~\ref{sec:Magnetic-Reconnection-Setup}. Likewise, the finite time integration results in side-lobes for signals (smaller peaks around the main signal), originating from the shape of the sinc-function. These side-lobes correspond to the sinc discussed in section~\ref{sec:Magnetic-Reconnection-Setup}. 
\begin{figure}[H]
    \centering
    \includegraphics[width=1.0\textwidth]{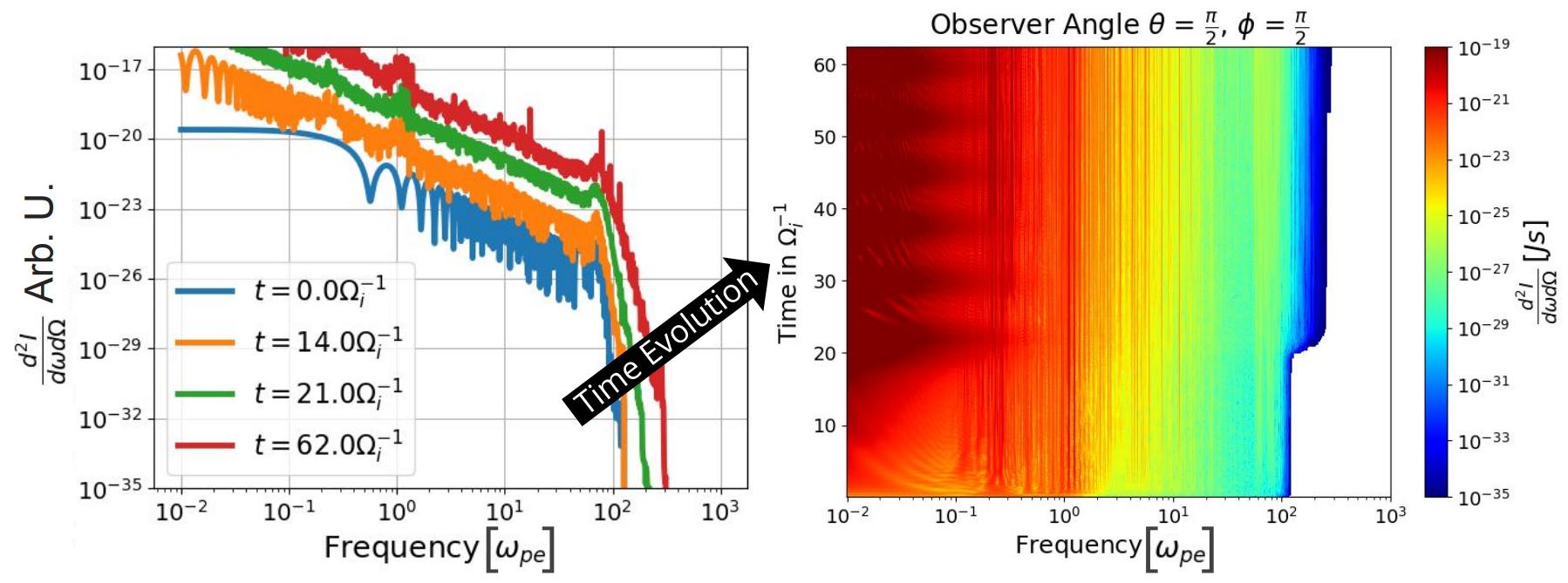}
    \caption{Transition from multiple plots of the time-integrated frequency spectrum to a 2D plot of the time-integrated frequency spectrum at each point in time. The $x$-axis denotes frequencies, the $y$-axis the time, and the color the intensity of the respective signal. The curves in the left figure are scaled by an arbitrary factor in order to properly show the time evolution of the curves, as they would otherwise overlap.}
    \label{fig:TimeEvoltionRad}
\end{figure}
Additionally, the emitted signal varies with the observation angle. Therefore, such plots are generated for all three orthogonal axes: the $x$-axis ($\theta=\frac{\pi}{2},\phi=0$) along the initial magnetic field lines, the $y$-axis ($\theta=\frac{\pi}{2},\phi=\frac{\pi}{2}$) along the density gradient, and the out-of-plane $z$-axis ($\theta=0$). For reference of the coordinate system definition, see figure~\ref{fig:Schematic}. It is sufficient to consider these three directions, as observations from directions in between show mixed contributions from each of the orthogonal directions respectively. The frequencies are given in multiples of the electron plasma frequency $\omega_{pe}$. Relevant radiative features are labeled by Roman letters from a) to d).
\begin{figure}[H]
    \centering
    \makebox[\textwidth]{\includegraphics[width=1.2\textwidth]{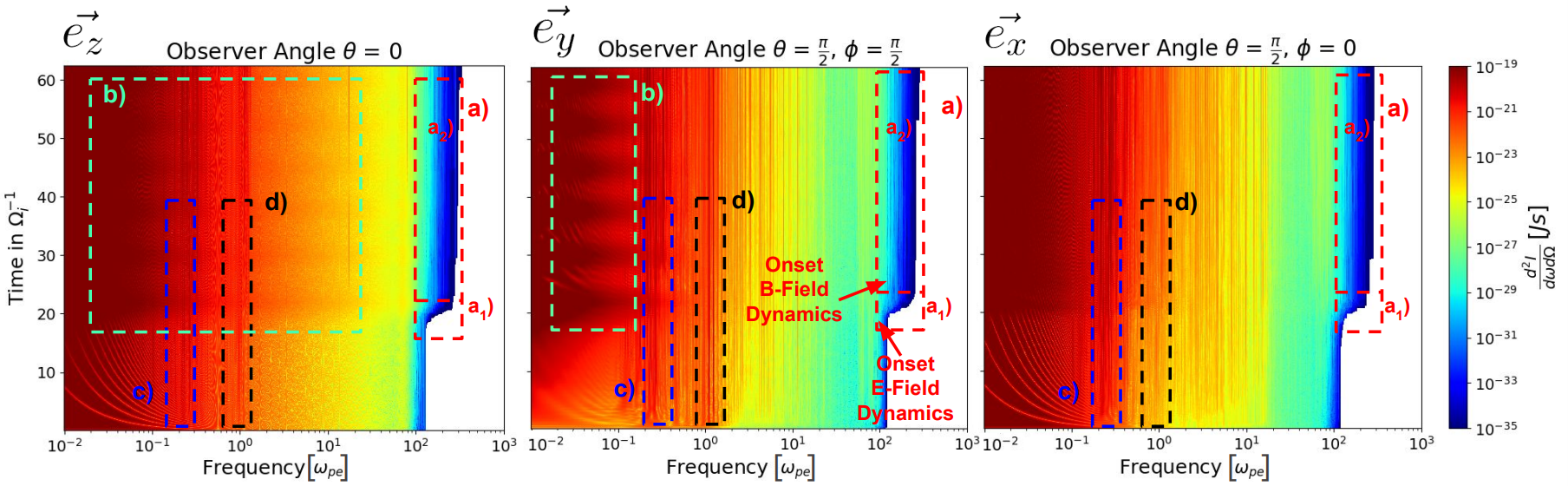}}
    \caption{Radiation over time in four different observation directions. Signals of great importance are discussed in the following sections. The roman letters represent: $a)$ radiation emitted by electrons undergoing a transition from electric $a_1)$ to magnetic field $a_2)$ dynamics. $b)$ denotes the radiation emitted by oscillating plasmoids. $c)$ is the electron gyrotron frequency. $d)$ is the plasma frequency.\\
    The figures only show radiation emitted in the three orthogonal coordinate axis. Any angle in between is found to have a mixture of the angles depicted here.} 
    \label{fig:RadOverTime}
\end{figure}
Plasma in an evolving magnetic field is expected to emit highly polarized radiation. To quantify the polarization in our radiation data, Stokes parameters are calculated~\cite{chandrasekhar1960radiative}, as shown in figure~\ref{fig:PolarizationOverTime}. The Stokes parameters—denoted as I, Q, U, and V—describe the polarization state. I is the total intensity, Q and U represent linear polarization along orthogonal and diagonal axes, respectively, and V measures circular polarization. Note that the figures shown here show the stokes parameters normalized to the total intensity I. Figure~\ref{fig:PolarizationOverTime} shows the most relevant polarization features, which are labeled by greek letters $\alpha$) to $\beta_2$).
Subsequently, we provide an overview of the relevant radiation features. An in-detail quantitative analysis is given in later sections. We divide them into three categories, based on relevance. First, we address the signals most likely to be observed in experiments and relevant for extracting experimental parameters. Second, we motivate a quantitative analysis of prospective signals that can be detected experimentally only in the highly relativistic case. This is because these prospective signals have a greater frequency than the plasma frequency only for extremely high temperatures, which is shown rigorously in section~\ref{sec:AnalyticModels}. Third, we sort out signals stemming from numerical effects.

\subsection{Experimentally Accessible Radiation Signatures}
%\todo{Vorverweis auf Bestaetigung(?) im analytischen Modell machen.}
%\subsubsection{Signal a)}
The radiation features discussed here resemble the dynamics shown in section~\ref{sec:Magnetic-Reconnection-Setup}. Consider the dynamics from the three previously defined stages: The initial dynamics, the electric field dominated dynamics and the magnetic field dominated dynamics. We note that the initial dynamics leave no relevant radiation features.\\
The electric field dominated dynamics, summarized in figure~\ref{fig:EZ}, is marked by radiation feature $a_1)$. The onset of this signature happens exactly at the time of the increase of out-of-plane electric field between 18~$\Omega_i^{-1}$, as indicated in the time evolution in figure~\ref{fig:EZ}. The increase in radiation energy lasts until approximately $22~\Omega_i^{-1}$, which matches the energy increase by the electrons indicated in figure~\ref{fig:PlasmoidOscillationEnergy}. The high frequency (above $10^2\omega_{pe}$) power-law radiation spectrum in figure~\ref{fig:TimeEvoltionRad} likely stems from the power law energy distribution of the final electrons after electric field acceleration as in figure~\ref{fig:PowerLaw}. In figure~\ref{fig:RadOverTime} this is indicated by signal $a_2)$. The transition from electric to magnetic field dynamics is represented in the radiation spectrum by signals $a_1)$ and $a_2)$, which are jointly summarized into signal $a)$.\\
The magnetic field dominated dynamics can also be seen in the polarization spectrum. The feature is denoted by $\alpha)$. Polarization is usually attributed to radiation from charged particles in a magnetic field, which indicates, that this signal stems from the electrons pushed into plasmoid structures (see figure~\ref{fig:PlasmoidOscillation}). Notably, observations from the $x$-direction show polarization along the $y$-axis, while those from the $y$-direction show polarization along the $x$-axis. This is expected, as each observation direction captures the projection of electron gyration onto its plane, resulting in the linear polarization observed. This also allows to infer information on the angle of observation $\theta$, since the polarization varies significantly in the frequency region above $10^2\omega_{pe}$ for that angle. This will become important in later derivations of section~\ref{sec:AnalyticModels}. The signal $\alpha)$ is subdivided into signals $\alpha_1)$ and $\alpha_2)$ indicating the time frame relevant for electric and magnetic field dynamics, analogous to the subdivisions of signal $a)$ into signal $a_1)$ and $a_2)$. The signal $\alpha_1)$ attributed to the electric field time frame is polarized as well, since the transition from electric to magnetic field dynamics is continuous. During the increase of the electric field, some electrons are already subject to magnetic field dynamics. This results in early polarization contributions to the high frequency spectrum.\\
\begin{SCfigure}[60][t!]
    \centering
    \includegraphics[width=0.5\textwidth]{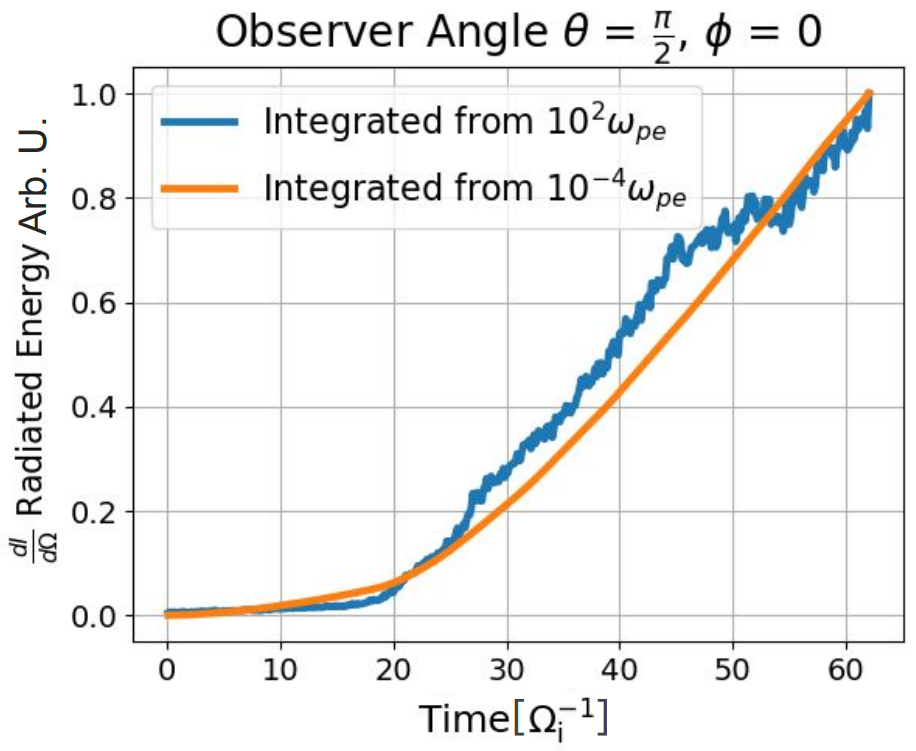}
    \caption{Since only radiation above the plasma frequency escapes the plasma, we desire to show that the frequency integrated radiated energy above the plasma frequency behaves similarly to the radiation emitted above $10^2\omega_{pe}$, a frequency region recognizable by its polarization signature. The radiated energy integrated over the full frequency spectrum is compared to the radiated energy integrated from $10^2\omega_{pe}$ onwards $\frac{dI}{d\Omega}=\int_{10^2\omega_{pe}}^\infty\frac{d^2I}{d\Omega d\omega}\,d\omega$. They show similar behavior, which justifies the analysis of this high frequency regions as a representative for the full frequency spectrum in the subsequent sections.}
    \label{fig:HighOmegLowOmeg}
\end{SCfigure}
These features resemble exactly the shift from electric field dynamics to magnetic field dynamics indicated by figures~\ref{fig:PSDensity} and~\ref{fig:PlasmoidOscillation}. The onset of the electric field dynamics is indicated by the rapid, nonlinear increase in radiated energy between 18~$\Omega_i^{-1}$ and 22~$\Omega_i^{-1}$, which corresponds to the rising energy of the electrons in figure~\ref{fig:PlasmoidOscillationEnergy}. This is especially evident when integrating the radiated energy over the entire spectral range. The transition to the magnetic field dynamics can be observed in the frequency integrated radiated energy. In figure~\ref{fig:HighOmegLowOmeg} the orange curve shows that past $22~\Omega_i^{-1}$ a linear increase in radiated energy is observed. This linear increase is an indicator for the magnetic field dynamics, since it stands in stark contrast to the rapid increase in radiated power between $18~\Omega_i^{-1}$ and $22~\Omega_i^{-1}$ when the electric field dynamics are relevant. However, an integral over all frequencies is not possible since the radiation below the plasma frequency does not escape the plasma. While an integral from the plasma frequency onwards $\frac{dI}{d\Omega}=\int_{\omega_{pe}}^\infty\frac{d^2I}{d\Omega d\omega}\,d\omega$ is possible in theory, in practice the plasma frequency of the reconnection event is not known. Instead we choose the highly polarized region past $10^2\omega_{pe}$ as a clear marker of the onset of the electric field and later the magnetic field dynamics. Figure~\ref{fig:HighOmegLowOmeg} shows that the radiation integrated from $10^2\omega_{pe}$ onwards is highly similar in behavior compared to the radiation integrated over the full frequency spectrum. Thus, the polarization is a strong experimental marker for the electric to magnetic field dynamics transition, which allows to uniquely identify the frequency region over which the spectral integral is supposed to be performed in order to obtain the curve in figure~\ref{fig:HighOmegLowOmeg}. This also allows for a clear separation between the time frames relevant for electric- and magnetic field dominated dynamics. Knowing these time frames becomes important later on, when extracting experimentally relevant parameters in section~\ref{sec:Verification}.\\
In subsequent sections, it will be shown that the knowledge about these two distinct time periods of electric field dominated dynamics and magnetic field dominated dynamics can be used to extract the ratio of out-of-plane electric field in the reconnection region, over plasmoid magnetic field. These are two essential quantities in examining magnetic field dynamics. This is further quantified in section~\ref{sec:AnalyticModels}.\\
Signal $b)$ characterizes plasmoid behavior, where accelerated electrons form plasmoids oscillating harmonically (see figure~\ref{fig:PlasmoidOscillation}), causing spectral energy to oscillate at the same frequency. As magnetic island structures, plasmoids emit highly polarized radiation, reflected in polarization feature $\beta_1)$ showing oscillations in polarization components. A similar feature, $\beta_2)$, exhibits oscillations from the y-direction but is likely suppressed in the x-direction due to noise. The earlier onset of $\beta_2)$ stems from electrons gyrating in the initial homogeneous magnetic field (left figure~\ref{fig:Schematic}). Both $\beta_1)$ and $\beta_2)$ represent magnetic field–dominated radiative dynamics, though $\beta_2)$ blends plasmoid dynamics with early magnetic field effects, distinguishing it from $\beta_1)$.
\subsection{Experimental Accessibility}
The feasibility of detecting these signals has to be studied specific to individual cases. As discussed in section~\ref{sec:Magnetic-Reconnection-Setup}, many quantities in this simulation are scaled down to achieve reasonable simulation times. Correspondingly, the number of electrons involved in a real magnetic reconnection event is much higher. In order to obtain an estimate for the expected radiated energy, we scale up all quantities accordingly.\\
We examine the feasibility for solar events, since the magnetic field was chosen accordingly. Based on this magnetic field we obtain a realistic number density of electrons. Thus, the plasma frequency in this setup is in the $10^2~\text{MHz}$ range. Solar events often involve length scales of $10^2\,\text{km}$~\cite{yamada_magnetic_2010}. Correspondingly, scaling the lowest energy signal of $10^{-8}~\text{Jy}$ up to the desired system size gives a signal strength of $10~\text{Jy}$ in the $10^2~\omega_{pe}=10~\text{GHz}$ range.\\
Such signals are resolvable by current ground based telescopes such as the \emph{Jansky} Very Large Array and observations with lower intensities have been made in the past~\cite{rodriguez_understanding_2023}. Nevertheless, these are rough estimates which require future simulations targeted towards specific reconnection scenarios.

\subsection{Prospective Radiation Features}
Notice that naively, we expect two spectral lines: We expect a spectral line stemming from the gyration of the electrons in the asymptotically constant initial magnetic field (see left figure~\ref{fig:Schematic}). We also expect a spectral line from the harmonic motion of particles around the zero crossing of the initial magnetic field.\\
The first of these signals is present in the form of a faint spectral line marked by $c)$. This corresponds to the frequency at which the electrons at the asymptotically constant magnetic field gyrate. However, signals below the plasma frequency do not escape the plasma and are therefore not measurable by real detectors. The zero crossing dynamics is shown in section~\ref{sec:AnalyticModels} to be far below the lowest limit of frequencies for which this simulation computes the radiation. For both signals to be above the plasma frequency and measured by a real detector the MR setup has to be highly relativistic as discussed in chapter~\ref{sec:AnalyticModels}.

\subsection{Spurious Signals from Numerical Effects}
The main numerical effect arises from the finite integration in time from equation~\ref{eq:radiation} resulting in side lobes, as discussed previously. Signal $d)$ shows a clear peak at the plasma frequency with side-lobes. While a peak around the plasma frequency is expected for a plasma in thermal equilibrium, the sudden disturbance by the perturbation term (equation~\ref{eq:Perturbation}) results in a sharp cutoff in the time integral and therefore in the side-lobes of the plasma frequency. This is an artificial effect due to the sudden disturbance by the perturbation term.
%\input{Sinc.tex}
%\todo{Textdopplungen beseitigen}

\section{Extracting Experimental Quantities from Radiation Data}\label{sec:AnalyticModels}
In the following, we build an analytic model linking signal $a)$ to particle dynamics. In section~\ref{sec:Magnetic-Reconnection-Setup} we argue that the electrons undergo a transition from electric field dominated dynamics to magnetic field dominated dynamics, which is resembled in the radiation spectra as shown in section~\ref{sec:Time-Evolution-of-Frequency-Spectra}. We use this to build a simplified, two step model. First, only radiation due to electric field acceleration, second the radiation from the magnetic field dynamics in the plasmoids is taken into account.

\subsection{Out-of-Plane Acceleration}
\label{sec:Reconnection-Flare-Acceleration}
First, the out-of-plane electric field is considered.
Figure~\ref{fig:PSDensity} shows a high electron density near the magnetic field's zero crossing, where particles are accelerated by the reconnection electric field. As the magnetic field is negligible in this region, its effects are omitted to isolate the electric field contribution.\\
Assuming an initial out-of-plane velocity, the radiated power per solid angle for $N$ electrons accelerated along their motion is given by~\cite{jackson1999classical}:
\begin{equation}\label{eq:radiationEq1}
    (\frac{dP}{d\Omega})_{E}=\frac{e^2 N}{16\pi^2\varepsilon_0c}\frac{\sin^2(\theta)}{(1-\beta\cos\theta)^5} \dot{\beta}^2
\end{equation}
Recall that all spatial coordinates and angles are indicated in figure~\ref{fig:Schematic}. We assume $\theta\neq 0$ since we expect no radiation emitted in this direction. Here, $e$, $N$, $\beta$ and $\dot{\beta}$ correspond to the electron charge, number of electrons accelerated by the electric field, electron velocity and electron acceleration respectively.\\ 
Up to first order, the evolution of the out-of-plane electric field can be approximated by a linear increase during the time of rapid amplitude growth $\Omega^{-1}t \approx 18-20$, as can be inferred from the linear fit in red from figure~\ref{fig:EZ}. The electric field per time is denoted by $\varepsilon$, and we write the equation of motion to solve for the acceleration, noting that by neglecting the magnetic field, the acceleration happens in the direction of motion. 
\begin{equation}\label{eq:Motion}
    \frac{d}{dt}(m_e c \gamma \beta) = \varepsilon e t \implies \gamma^3\dot{\beta}=\frac{\varepsilon e t}{m_ec}
\end{equation}
%In the considered scenario, the non-relativistic initial drift velocity $|\vec \beta_0|$ allows for the approximation $|\vec \beta_0|\approx 0$. Solving for $\beta$ gives the following expression.
%\begin{equation}
%    \beta(t) = \frac{e \varepsilon t^2}{\sqrt{4 m_e^2 c^2 + e^2 \varepsilon^2 t^4}}
%\end{equation}
Plugging the expression for $\dot{\beta}$ into the radiation equation~\ref{eq:radiationEq1} yields the following formula.
\begin{equation}\label{eq:recRateFinal}
    (\frac{dP}{d\Omega})_{E} = \frac{e^4N\varepsilon^2t^2}{16\pi^2\varepsilon_0c^3m_e^2}\frac{\sin^2(\theta)}{(1-\beta\cos(\theta))^5\gamma^3}
\end{equation}
For small values of $\beta$, in the non-relativistic regime, the radiation formula simplifies significantly and scales quadratically with the electric field $(\frac{dP}{d\Omega})_{E}\propto \varepsilon^2t^2=E^2$. Figure~\ref{fig:PowerLaw} shows that electrons span a wide range of energies, up to $\sim 200$~keV. Since $m_e=511~\text{keV/c}$, we find $\gamma\sim 1$, which allows to neglect relativistic effects. 
%Note that all terms with $\frac{e \varepsilon t^2}{m_e c} \ll 1$ may be neglected. This is because $e \varepsilon t^2$ is proportional to the momentum, as seen in equation~\ref{eq:Motion} via $\frac{dp}{dt} = \varepsilon e t$. From the relativistic energy-momentum relation $\gamma^2 m_e^2 c^4 = p^2 c^2 + m_e^2 c^4$, we find:
%\[
%\frac{e \varepsilon t^2}{m_e c} = 2 \frac{pc}{m_e c^2} = 2 \sqrt{\gamma^2 - 1}
%\]
%From the energy distribution in figure~\ref{fig:PlasmoidOscillationEnergy}, we infer that with 100~keV kinetic energy, $\gamma$ is on the order of 1 here, which justifies $\frac{e \varepsilon t^2}{m_e c} \ll 1$. Therefore, we approximate:
%\begin{equation}\label{eq:recrateFinal}
%    \left( \frac{dP}{d\Omega} \right)_{\text{rec}} = \frac{\mu_0 e^4 m_e^2 \varepsilon^2 c N}{\pi^2} \sin^2(\Theta) t^2
%\end{equation}
% To get the Energy over Time:
% \begin{equation}\label{eq:RecE}
%     \left( \frac{dE}{d\Omega} \right)_{\text{rec}} = \frac{\mu_0 e^4 m_e^2 \varepsilon^2 c N}{3 \pi^2} \sin^2(\Theta) t^3
% \end{equation}

\subsection{Synchrotron Power Law Radiation}
\label{sec:Synchrotron Power Law Radiation}
In the next step, the electrons in plasmoids are considered, which exhibit a power-law energy spectrum, displayed in figure~\ref{fig:PowerLaw}. As discussed in section~\ref{sec:MagneticFieldDynamics} the electrons undergo dominantly magnetic field dynamics. Therefore, the emitted radiative power per electron can be computed by considering the case of a force acting perpendicular to the direction of motion~\cite{jackson1999classical}.
\begin{equation}\label{eq:radiationEq2}
    (\frac{dP}{d\Omega})_{E}=\frac{e^2}{16\pi^2\varepsilon_0c}\frac{1}{(1-\beta\cos(\theta))^3}\left ( 1-\frac{\sin^2(\theta)\cos^2(\phi)}{\gamma^2(1-\beta\cos(\theta))^2}\right ) \dot{\beta}^2
\end{equation}
Furthermore, considering the charge density in figure~\ref{fig:PlasmoidOscillation} reveals that the electrons gather narrowly in a region of approximately constant magnetic field. The same plot shows that past 20~$\Omega_i^{-1}$ the electrons experience no significant change in magnetic field, as the magnetic field line density remains the same.\\
The $N$ electrons accelerated by the electric field follow a power law energy distribution $f(\gamma)$ with a normalization constant $f_0$ and the parameter $\alpha$.
\begin{equation}
    f(\gamma)=Nf_0\gamma^{-\alpha}
\end{equation}
We assume that the particles span over all energies such that $\gamma_0=1$ and $\gamma_1=\infty$. This gives the normalization constant $f_0=\alpha-1$. To evaluate the integral, the acceleration of the particle is written in terms of the absolute magnetic field inside the plasmoid indicated in figure~\ref{fig:PlasmoidOscillation}. Thus we approximate that all radiating electrons experience the same magnetic field.
%This magnetic field is on the order of $B_0\approx 0.5B_0$ for most electrons as discussed in section~\ref{sec:Magnetic-Reconnection-Setup} and depicted by the bottom figure~\ref{fig:PlasmoidOscillation}.
\begin{equation}
    \dot{\beta}=\frac{eB}{m_e\gamma}\beta
\end{equation}
The full integral in terms of $\gamma$ and $\beta$ is given by the following expression.
\begin{equation}\label{eq:radSyncFinal}
    \left(\frac{dP}{d\Omega}\right)_{B} = \frac{e^4B^2N(\alpha-1)}{16\pi^2\varepsilon_0 cm_e^2} 
    \underbrace{\int_1^\infty \frac{\gamma^{-(\alpha+2)}}{(1-\beta\cos(\theta))^3}\left ( 1-\frac{\sin^2(\theta)\cos^2(\phi)}{\gamma^2(1-\beta\cos(\theta))^2}\right ) \beta^2
         \,d\gamma
    }_{\psi_\alpha(\theta,\phi)}
\end{equation}
We denote the integral as $\psi_\alpha(\theta,\phi)$, which only depends on $\alpha,\theta$ and $\phi$. The plasmoid structures show an approximately circular symmetry as depicted in figure~\ref{fig:PlasmoidOscillation}. Therefore we replace the dependence on $\phi$ in $\psi_\alpha(\theta,\phi)$ with the average over the polar angle $\langle\psi_\alpha(\theta)\rangle=\frac{1}{2\pi}\int_0^{2\pi}\psi_\alpha(\theta,\phi)\,d\phi$.\\
Taking the ratio of the emitted powers from equation~\ref{eq:recRateFinal} and equation~\ref{eq:radSyncFinal} reveals a relationship between the out-of-plane reconnection electric field and the magnetic field inside the plasmoid.
\begin{equation}\label{eq:final}
    (\frac{E}{cB})^2=\langle \psi_\alpha(\theta)\rangle_\phi (\alpha-1)\frac{(dP/d\Omega)_{E}}{(dP/d\Omega)_{B}}
\end{equation}
\begin{SCfigure}[50][t!]
    \centering
    \includegraphics[width=0.5\linewidth]{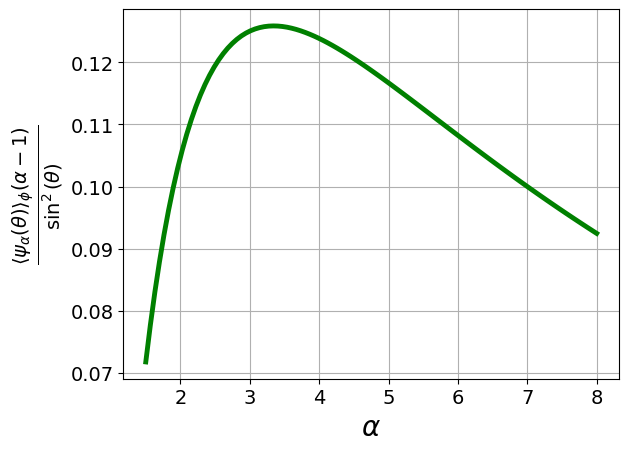}
    \caption{The proportionality constant in equation~\ref{eq:final} with respect to $\alpha$ is displayed. It clearly shows that the choice of $\alpha$ does not influence the predicted order of magnitude. For higher, physically irrelevant $\alpha$, the order of magnitude of $\frac{\langle \psi_\alpha(\theta)\rangle(\alpha-1)}{\sin^2(\theta)}$ decreases.}
    \label{fig:ALPHA}
\end{SCfigure}
The remaining two free parameters $\theta$ and $\alpha$ can be determined. The polarization depicted in figure~\ref{fig:PolarizationOverTime} varies with the direction of observation and thereby, $\theta$ can be inferred. The final free parameter $\alpha$ is known to be on the order of magnitude of $\sim 1$. Additionally, when radiation is accompanied by a particle shower, $\alpha$ can be inferred directly from the particle shower. The order of magnitude of $(\frac{E}{cB})^2$ is independent of the choice of $\alpha$, since the proportionality constant from figure~\ref{fig:ALPHA} does not change the order of magnitude with $\alpha$.\\
Formula~\ref{eq:final} is a key result, enabling order-of-magnitude estimates of the reconnection electric field to plasmoid magnetic field ratio using radiation data alone. This ratio has direct physical meaning as the energy density ratio $\frac{\frac{1}{2}\varepsilon_0E^2}{\frac{1}{2\mu_0}B^2} = \left(\frac{E}{cB}\right)^2$. Its importance is discussed in Section~\ref{sec:Discussion}, and the following sections provide numerical verification and demonstrate its practical extraction.

\subsection{Predicting E/B from Radiation Data}
\label{sec:Verification}
\begin{figure}[H]
    \centering
    \includegraphics[width=0.99\textwidth]{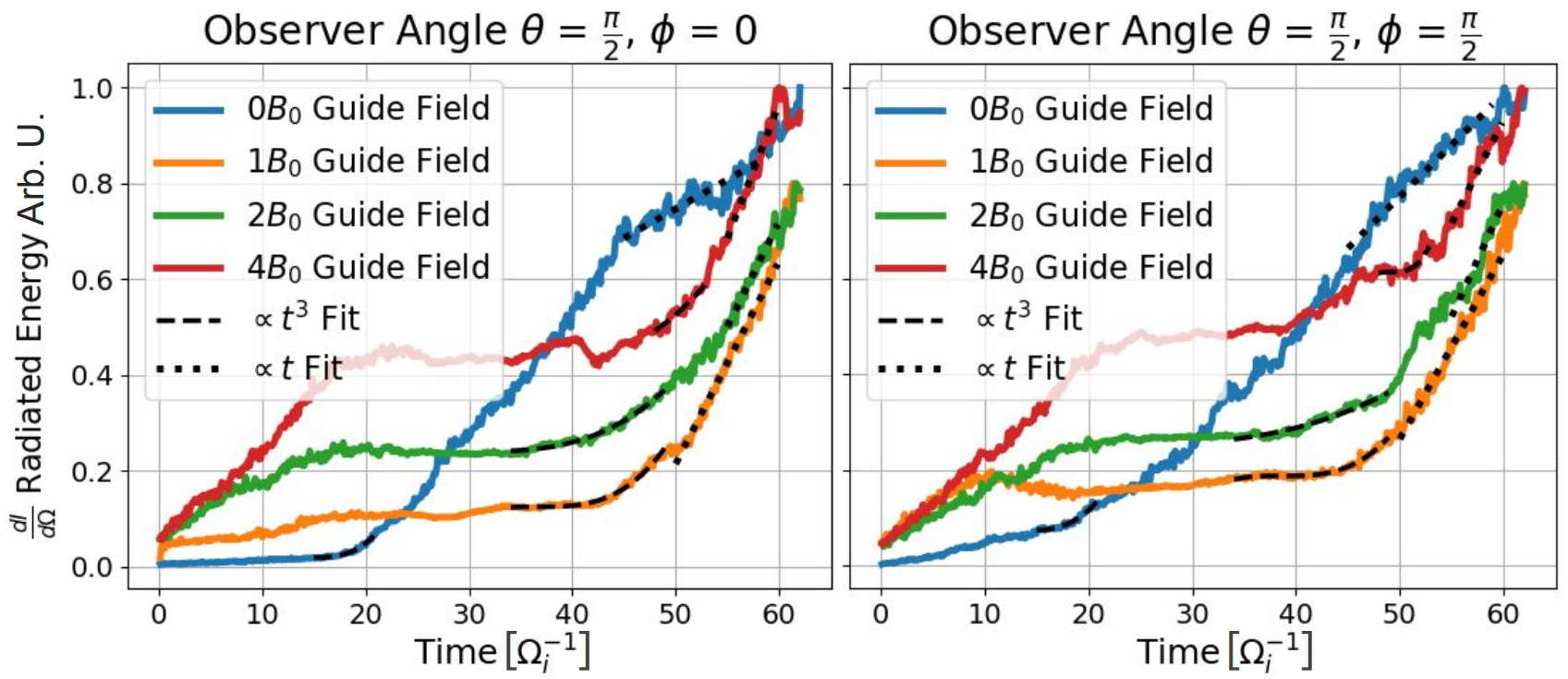}
    \caption{Frequency integrated radiated energy. (Left) Observation along the magnetic field lines. (Right) Observations along the density gradient of the plasma. In both cases the radiation for each guide field is scaled by an arbitrary factor for better visibility of the data. The radiated energy is correspondingly depicted in arbitrary units.}
    \label{fig:proptoFit}
\end{figure}

\begin{figure}[H]
    \centering
    \makebox[\textwidth]{\includegraphics[width=1.2\textwidth]{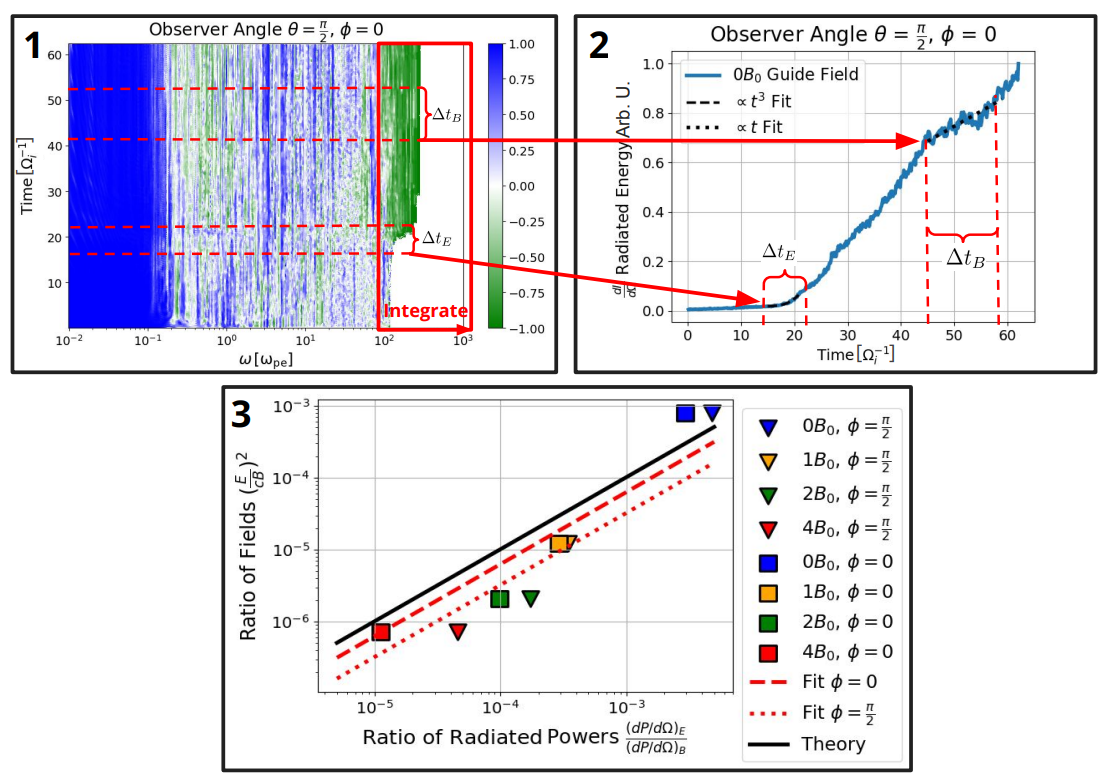}}
    \caption{The three-step schematic illustration of the analysis necessary to obtain an order of magnitude estimate of $E/B$. The third figure summarizes the results for all guide-field simulations, validating the theoretical prediction with accuracy within an order of magnitude.}
    \label{fig:JUHUU}
\end{figure}

We verify the central results obtained from theoretical considerations of the previous sections~\ref{sec:Reconnection-Flare-Acceleration} and~\ref{sec:Synchrotron Power Law Radiation}. Three central claims must be verified.
\begin{itemize}
    \item The radiated power during the electric field dominant dynamics scales as $(\frac{dP}{d\Omega})_E\propto t^2$ in time (see equation~\ref{eq:radiationEq1}). Correspondingly, the radiated energy scales as $(\frac{dI}{d\Omega})_E\propto t^3$ in time. 
    \item The radiated power during the magnetic field dominant dynamics scales as $(\frac{dP}{d\Omega})_B\propto \text{const}$ in time (see equation~\ref{eq:radiationEq2}). Correspondingly, the radiated energy scales as $(\frac{dI}{d\Omega})_B\propto t$ in time.
    \item The ratio of radiated energies allows to obtain the ratio of reconnection-electric-field to plasmoid magnetic field $(\frac{E}{cB})^2\propto\frac{(dP/d\Omega)_E}{(dP/d\Omega)_B}$ (see equation~\ref{eq:final}).
\end{itemize}
With these central claims established, we proceed to verify them. Since the radiation plugin outputs radiated energy rather than power, we test the corresponding radiation energy scalings across four simulations to assess robustness. The first uses the baseline setup; the others introduce guide fields $B_G = 1B_0$, $2B_0$, and $4B_0$ to systematically vary the reconnection electric field $E$. These same simulations are later used to test the extraction of $\left(\frac{E}{cB}\right)^2$, as guide fields are known to modify $E$ in a controlled and predictable manner~\cite{fu_tripolar_2018}.\\
However, the introduction of a guide field leads to additional effects. High energy electrons are accelerated later in time by the electric field. Thus, the $\propto t^3$ scaling appears later in time for the guide field simulations compared to the simulation with no guide field. This is a well known effect~\cite{GuideFieldLater} and its detailed examination is out of the scope of this work. The time frames corresponding to the electric- and magnetic field dynamics are listed in the following for all simulations.
\begin{table}[H]
    \centering
    \begin{tabular}{|c|c|c|c|c|c|}
         \hline
         Timeframe & $B_{\text{Guide}}=0B_0$ & $B_{\text{Guide}}=1\cdot B_0$ & $B_{\text{Guide}}=2\cdot B_0$ & $B_{\text{Guide}}=4\cdot B_0$ \\
         \hline
         $\Delta t_E$ & $18-22~\Omega_i^{-1}$ & $34-49~\Omega_i^{-1}$ & $34-49~\Omega_i^{-1}$ & $48-52~\Omega_i^{-1}$\\
         $\Delta t_B$ & $45-59~\Omega_i^{-1}$ & $50-60~\Omega_i^{-1}$ & $50-60~\Omega_i^{-1}$ & $50-60~\Omega_i^{-1}$\\
         \hline
    \end{tabular}
    %\caption{Caption}
    \label{tab:placeholder}
\end{table}
Figure~\ref{fig:proptoFit} shows the corresponding scaling for the electric- and magnetic field time frames. We choose $\theta=\frac{\pi}{2}$ and verify for both $\phi=0,\frac{\pi}{2}$. The angle $\theta=\frac{\pi}{2}$ is most relevant since this is the direction from which particle showers are expected to be observed on earth, which indicate a magnetic reconnection event. We see that the scaling of both $\propto t$ and $\propto t^3$ occurs in all cases. The level to which the scaling is pronounced varies with guide field strength and observation directions. Nevertheless, we find good agreement with the predicted scaling, especially for the guide field free case. 
Finally, we show that we can extract the $E/B$ ratio. The procedure necessary for this, is shown by the schematic in figure~\ref{fig:JUHUU}. The extraction of $E/B$ is achieved in a three step process.
\begin{enumerate}
    \item Identify the polarized frequency region as indicated by $\alpha )$ in figure~\ref{fig:PolarizationOverTime}. Integrate the radiated energy starting from the frequency where the polarization of the signal begins $\frac{dI}{d\Omega}=\int_{10^2\omega_{pe}}^\infty\,\frac{d^2I}{d\Omega d\omega}\,d\omega$ (area inside the red box in figure~\ref{fig:JUHUU} number 1. This gives the figure in step 2.\\
    Additionally, the time frame for electric and magnetic field dominated dynamics ($\Delta t_E$ and $\Delta t_B$) can be extracted. This is done by considering the increase in radiated energy at high frequencies for the electric field dynamics. For the magnetic field dynamics any point in time after the electric field dominated dynamics time frame can be chosen. This is because the radiated power for the magnetic field dominated dynamics in equation~\ref{eq:final} is constant over time.
    \item The second step shows the frequency integrated radiated energy over time. We identify the increase by $\propto t_E^3$ for the electric field dynamics and $\propto t_B$ for the magnetic field dynamics. The subscripts $E$ and $B$ are given to the times to highlight that these are two different time frames. We fit the corresponding functions $C_1\cdot t_E^3$ and $C_2\cdot t_B$ in the two respective time frames, with $C_1$ and $C_2$ constants. This allows to compute the ratio $\frac{(dP/d\Omega)_E}{(dP/d\Omega)_B}$ in step 3.
    \item Since the fits in step 2 were performed on the radiated energy instead of power, coefficient obtained from the fit needs to be adjusted. If $I=C_1\cdot t_E^3$ then $P=3\cdot C_1\cdot t_E^2$ and correspondingly if $I=C_2\cdot t_B$ then $P=C_2$. This allows to compute $(\frac{E}{cB})^2=\frac{\langle\psi_{\alpha}(\theta)\rangle_\phi (\alpha-1)}{\sin^2(\theta)}\cdot\frac{3\cdot C_1\cdot t_E^2}{C_2}$ (see equation~\ref{eq:final}). The quantity $\alpha\sim 10^0$ has little effect on the order of magnitude estimate as indicated in figure~\ref{fig:ALPHA} and the quantity $\theta$ can be estimated form the polarization data in figure~\ref{fig:PolarizationOverTime} as discussed previously. Notice that a dependence on time remains, which reflects the fact that the electric field changes in time as well. This is illustrated in figure~\ref{fig:EZ}. Additionally, notice that $t_E$ is a point in time relative to the start of the electric field dynamics. Thus, no knowledge on the start of the reconnection event is required. For the verification here, $t_E$ is chosen where $E$ is maximal.\\
    In figure~\ref{fig:JUHUU} number~3 the ratio of radiated powers is obtained as above for all simulations, observing from two different directions $\phi=0,\frac{\pi}{2}$. The electric to magnetic field can be extracted from the simulation, giving the values indicated by the squares and triangles. The theoretical predictions for these values from equation~\ref{eq:final} are given by the black line. 
\end{enumerate}
The third figure shows good agreement for an order of magnitude estimate of the ratio $(\frac{E}{cB})^2$ from the ratio of radiated powers. Moreover, it shows that the observations from $\phi=0$ and $\phi=\frac{\pi}{2}$ within an order of magnitude, which allows to take the average over $\phi$ as discussed previously. Thus, all free parameters in the proportionality factor $\frac{\langle \psi_\alpha(\theta)\rangle (\alpha - 1)}{\sin^2(\theta)}$ are eliminated.\\
While the choice of time frame for the magnetic field–dominated regime introduces some ambiguity, this is mitigated by the fact that any period after the electric field dynamics ($22~\Omega_i^{-1}$) yields estimates of $(\frac{E}{cB})^2$ within the same order of magnitude. Due to inherent noise in the radiation data, fitting windows must be adjusted accordingly. For experimental contexts, where exact reconnection timescales may be unknown, robustness is essential. Empirically, variations in the selected time frame lead to deviations of at most one order of magnitude—primarily in the zero and high guide field cases—while intermediate cases remain stable. Given that the simulations span three orders of magnitude in $(\frac{E}{cB})^2$, the method demonstrates reliable accuracy. In principle, since $(\frac{E}{cB})^2$ can be made arbitrarily small by increasing the guide field, and was observed on the order of $10^{-1}$~\cite{pritchett2001gem}, a one-order deviation remains within acceptable bounds.\\  
This verifies the central result of the paper and allows for the estimate of the central quantities $E$ and $B$. The applications of this result are discussed in section~\ref{sec:Discussion}. 

\subsection{Prospective Features in Highly Relativistic Reconnection}
The subsequent section provides an outlook on further quantities which can be extracted from radiation data in highly relativistic magnetic reconnection cases. These can be performed in future works. 
\subsubsection{Electron Gyration Frequency}
\label{sec:Gyro}
In figure~\ref{fig:RadOverTime}, label b) marks the gyration frequency of the radiation electrons in the asymptotic magnetic field $B_0$ of the initial magnetic field configuration indicated by figure~\ref{fig:Schematic}. The signal labeled by b) can be linked to the gyration frequency $\omega_{g}$ by computing it from the parameters introduced in section~\ref{sec:Magnetic-Reconnection-Setup}.\\
In our setup, the frequency would not be observable experimentally, as it is smaller than the plasma frequency. However, knowing this frequency allows to extract important parameters via $\omega_{g}=\frac{eB}{\gamma m_e}$ such as the initial drift Lorentz factor of the electrons or asymptotic magnetic field. In the following a limit is derived, in which the gyration frequency is above the plasma frequency and therefore observable.\\
For this signal to be observed, we require it to be greater than the plasma frequency $\omega_g\geq\omega_{pe}$. Considering the expression for the two frequencies in the relativistic case $\omega_g=\frac{eB}{\gamma m_e}\geq \omega_{pe}=\sqrt{\frac{n_0e^2}{\varepsilon_0\gamma m_e}}$, we would like to eliminate the magnetic field by using the equilibrium condition derived by Harris~\cite{harris1962plasma} and Hoh for the relativistic case~\cite{hoh1966stability}. 
\begin{equation}\label{eq:StabilityConditions}
    B_0^2 = 2\mu_0k_B(T_i+T_e)n_0    
\end{equation}
Inserting the condition in the inequality required, a constraint on the temperature of the system is found.
\begin{equation}\label{eq:intermediate}
    2k_B(T_i+T_e)\geq\gamma m_ec^2
\end{equation}
Note that the left hand side corresponds to the total energy of the electrons. Thus, such signals can only be observed in the limit of highly relativistic temperatures. Additionally, the equipartition theorem for such temperatures can be considered, giving $\langle E\rangle=3k_BT_e$~\cite{equipartition}. We thereby assume local thermal equilibrium for the Harris steady state solution. Thus, in the highly relativistic case, the result can be expressed as a limit on the temperature ratio.
\begin{equation}\label{eq:TiTe05}
    \frac{T_i}{T_e}\geq\frac{1}{2}
\end{equation}
These limits motivate further work on radiation analysis in highly relativistic magnetic reconnection setups, as distinct spectral lines can be observed in these scenarios, allowing for the direct extraction of relevant quantities like the magnetic field.
\subsubsection{Observation of the Zero Crossing Dynamics}
\label{sec:Zero}
\begin{SCfigure}[50][h!]
    \centering
    \includegraphics[width=0.55\linewidth]{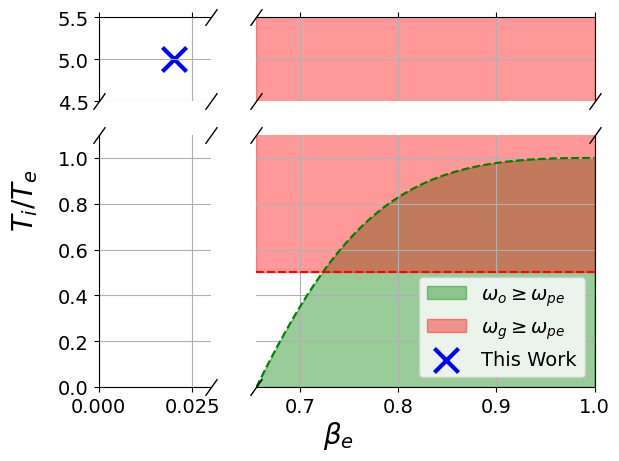}
    \caption{Visualization of the constraints on the temperature ratio $T_i/T_e$ necessary for the detection of the frequency caused by the gyration of electrons in the initial magnetic field (red) and the zero crossing frequency (green). Notice that below $\beta_e\approx 0.65$ the signal $\omega_o$ can never be observed. Notice also that below $\beta_e=0.65$ the limit $\omega_g\geq\omega_{pe}$ is no longer marked in red. This is because we additionally required an highly relativistic velocity in equation~\ref{eq:intermediate} for $\omega_g$ to be observed. Since $\beta_e=0.02$ is not highly relativistic, we do not mark this area in red.}
    \label{fig:TiTeLimit}
\end{SCfigure}

Another signal which we expect to observe is the signal originating from periodic movement of particles around the magnetic field zero crossing in the initial magnetic field configuration shown in figure~\ref{fig:Schematic}. That is, the electrons that move close to the zero crossing of the magnetic field, experience a small force, increasing with distance to the origin (see magnetic field in equation~\ref{eq:Harris}), pushing them back to the origin, which mimics harmonic oscillator behavior.\\
In our setup, the frequency of the signal associated with this periodic movement is lower than the computed minimum radiation frequency in the PIC simulation and thus lower than the plasma frequency. However, observing this signal allows for deducing quantities like the current sheet half width $\lambda$, as given by the equation~\ref{eq:omegao}. This makes a limit on the observability of the zero crossing frequency highly interesting.\\
Considering the electrons around the zero crossing of the magnetic field, it is expected that for small initial displacements, the Lorentz force acts approximately linear, resulting in harmonic oscillations. The expression for the magnetic field is linearized from equation~\ref{eq:Harris} setting $\text{tanh}(\frac{y}{\lambda})\approx\frac{y}{\lambda}$. The equation of motion gives that of an harmonic oscillator as expected, for an electron moving with velocity $\beta_e$ and corresponding Lorentz factor $\gamma_e$.
\begin{equation}\label{eq:omegao}
    \ddot{y}\approx -\frac{eB_0\gamma_e\beta_e c}{\lambda m_e}y
\end{equation}
Here, the oscillation frequency is given by $\omega_o^2=\frac{B_0 e\gamma_e\beta_e c}{\lambda m_e}$. For a signal to be observed experimentally, its frequency must be greater than that of the plasma frequency and we employ the condition $\omega_0\geq\omega_{pe}$ again. Likewise, to eliminate $B_0$ and $\lambda$ we employ the Harris stability condition from equation~\ref{eq:StabilityConditions} and in addition, the stability condition relating the parameter $\lambda$ for a stable solution of the Harris current sheet model~\cite{harris1962plasma}.
\begin{equation}\label{eq:HarrisMe}
    \frac{B_0}{\lambda}=n_0e\mu_0c(\beta_e\gamma_e+\beta_i\gamma_i)
\end{equation}
This results in a constraint on the initial electron and ion velocity.
\begin{equation}    \beta_e\gamma_e^2(\gamma_e\beta_e+\gamma_i\beta_i)\geq1
\end{equation}
Using the relation $\beta_i/\beta_e=T_i/T_e$, we find a constraint on the temperature ratio by equation~\ref{eq:TiTeBe}. Both constraint~\ref{eq:TiTeBe} and constraint~\ref{eq:TiTe05} is displayed in figure~\ref{fig:TiTeLimit}. 
\begin{equation}\label{eq:TiTeBe}
    \frac{\gamma_e^3-\gamma_e-1}{\gamma_e^2-1+\beta_e^2(\gamma_e^3-\gamma_e-1)}\geq\frac{T_i}{T_e}
\end{equation}
Likewise the highly relativistic reconnection case becomes of interest, since equation~\ref{eq:TiTeBe} allows only for solutions above $\beta_e\approx 0.65$. The limit on the signal $\omega_g$ of the electrons in the initial magnetic field is also indicated. Most notably, both signals can only be observed simultaneously in a well defined region of $T_i/T_e$ combinations. This is only possible for $\beta_e\geq0.72$.\\
These results provide a promising outlook for the highly relativistic case and show that the detection of both the gyration frequency and the zero crossing frequency narrow the parameter range of the observed magnetic reconnection scenario considerably.

\section{Discussion and Conclusion}
\label{sec:Discussion}
This work introduces a new framework for diagnosing magnetic reconnection remotely through radiation alone, offering a novel, data-driven tool to probe plasma conditions in extreme astrophysical environments. Our study identifies signals fundamental signals that are linked to the characteristic magnetic reconnection dynamics, which include: 
\begin{itemize}
    \item A rise in highly polarized radiation at high frequencies during the onset of magnetic reconnection, driven by electric field acceleration and subsequent trapping in magnetic islands. These signals are labeled a) and $\alpha$) in Figures~\ref{fig:RadOverTime} and~\ref{fig:PolarizationOverTime}.
    \item Periodic modulation of radiated energy and strong linear polarization at lower frequencies, arising from plasmoid oscillations. These features are labeled b), $\beta_1$), and $\beta_2$) in Figures~\ref{fig:RadOverTime} and~\ref{fig:PolarizationOverTime}.
\end{itemize}
A central achievement of this study is the ability to extract the ratio $(\frac{E}{cB})^2$ directly from radiative signals—transforming a traditionally unmeasurable quantity into an observable. This enables a new class of remote diagnostics for reconnection physics. The qualitative features named above aid with the extraction of the reconnection-electric-field over magnetic field ratio. Most noticeably, the simulated data is in good agreement with theoretical prediction, allowing for direct estimates of the $E/B$ ratio. Even with significant changes in guide field strength, the method remains accurate within an order of magnitude. Likewise, the method proves robust to variable choice of timescales considered for the analysis. This is sufficient for distinguishing physical regimes in astrophysics. Additionally, we show that many more quantities can be inferred from radiation signals, when highly relativistic magnetic reconnection is examined.\\ 
We highlight the relevance of the extraction of the reconnection electric field to plasmoid magnetic field ratio $E/B$. It allows to examine remote, previously inaccessible magnetic reconnection events. The electric and magnetic field are key observables in examining problems such as the contribution to magnetic reconnection of anomalous resistivity due to wave-particle interaction. The observables also provide clarification on the role of electric fields in formation of power law spectra and plasmoid dynamics in general via their magnetic fields~\cite{OutstandingQuestions}. Moreover, estimates of this ratio can provide an insight on wether magnetic reconnection is viable as a mechanism for observed astrophysical phenomena.\\
Likewise, our qualitative observations allow to illuminate the role of magnetic reconnection in many phenomena when radiation data is available and aid with the identification of magnetic reconnection. Additionally, our examinations significantly extend the observations by previous works~\cite{zhang2018large, zhang2020radiationI, zhang2022radiationII}, which identified the polarization swing as a unique feature of magnetic reconnection. We provide a more holistic view on the polarization swing, since we show it appears in a well defined frequency region above $10^2\omega_{pe}$. Additionally, we find a distinct polarization swing created by the oscillation of plasmoids.\\
Most significantly, this work opens a new research direction in fields adjacent to plasma- and astrophysics. Our work lays the theoretical foundation for directions which can be expanded on in the future by using the PIConGPU radiation plugin to examine the radiation of highly relativistic magnetic reconnection. In section~\ref{sec:AnalyticModels} we provide insights on promising signals in future, highly relativistic simulations. Additionally, the scaling of the $E/B$ ratio seeks for experimental verification. Limits on observability must be set, like examining the effect of additional dynamics introduced e.g. 3D setups~\cite{3D1, 3D2} or turbulences~\cite{turbulent1, turbulent2}. Future works can additionally examine the established signals in contexts of magnetic reconnection simulations tailored more towards specific physical events like geomagnetic storms, events in the solar-environment or high energy astrophysical phenomena. All of this provides an outlook for an exciting and promising research direction.

\section*{Acknowledgments}
S.~K.~E.\ highly acknowledges the financial support of the Lions Club Neu-Isenburg and the German Scholarship Foundation. K.~S.\ and M.~B.\ are funded by the Center for Advanced Systems Understanding (CASUS) which is financed by Germany’s Federal Ministry of Education and Research (BMBF) and by the Saxon Ministry for Science, Culture and Tourism (SMWK) with tax funds on the basis of the budget approved by the Saxon State Parliament. This research used the Hemera computing cluster at the Helmholtz-Zentrum Dresden -- Rossendorf e.\,V.

\bibliography{references} 

\end{document}